\documentclass[aip,cha,twocolumn,reprint]{revtex4-1}
\usepackage[pdftex]{graphicx}
\usepackage[usenames,dvipsnames,svgnames,table]{xcolor}

\usepackage{mathtools}
\usepackage{amsmath,amssymb}
\usepackage{comment}
\usepackage[caption=false]{subfig}
\usepackage{tabularx}

\newcommand{\arXiv}[1]{{\tt \href{http://arXiv.org/abs/#1}{arXiv:#1}}}
\newcommand{\arxiv}[1]{{\tt \href{http://arXiv.org/abs/#1}{arXiv:#1}}}

\newcommand{\ii}{\mathrm{i}}
\newcommand{\dd}{\mathrm{d}}
\newcommand{\tr}{\mathrm{tr}}

\renewcommand{\vec}[1]{\mathbf{#1}}

\definecolor{black}{rgb}{0,0,0}
\definecolor{myblue}{rgb}{0,0,0.6}
\definecolor{myred}{rgb}{0.6,0,0}
\newcommand{\cmedit}[1]{\textcolor{black}{#1}}
\newcommand{\rgedit}[1]{\textcolor{black}{#1}}

\graphicspath{{./figures/}}

\begin{document}

\title{Adjoint eigenfunctions of temporally recurrent single-spiral solutions in a simple model of atrial fibrillation.}

\author{Christopher D. Marcotte}
\affiliation{School of Physics, Georgia Institute of Technology}
\author{Roman O. Grigoriev}
\affiliation{School of Physics, Georgia Institute of Technology}

\date{\today}

\begin{abstract}
This paper introduces a numerical method for computing the spectrum of adjoint (left) eigenfunctions of spiral wave solutions to reaction-diffusion systems in arbitrary geometries. 
The method is illustrated by computing over a hundred eigenfunctions associated with an unstable time-periodic single-spiral solution of the Karma model on a square domain. 
We show that all leading adjoint eigenfunctions are exponentially localized in the vicinity of the spiral tip, although the marginal modes (response functions) demonstrate the strongest localization. 
We also discuss the implications of the localization for the dynamics and control of unstable spiral waves. 
In particular, the interaction with no-flux boundaries leads to a drift of spiral waves which can be understood with the help of the response functions.
\end{abstract}


\maketitle

\begin{quotation}
Fibrillation is a complex cardiac rhythm featuring multiple spiral waves that keep breaking up and merging. Some features of these complex dynamics can be understood by focusing on the dynamics of individual spirals and their interaction with neighboring spirals. While the dynamics of isolated spirals has been studied rather extensively, \cmedit{a description of the interaction between spiral waves is significantly less developed}. 
The adjoint eigenfunctions associated with isolated spiral wave solutions can provide such a description. This paper explains how these adjoint eigenfunctions can be computed for spiral waves of arbitrary size and temporal complexity and discusses what the structure of adjoint eigenfunctions tells us about the interaction of spiral waves with their surroundings.
\end{quotation}

\section{Introduction}

Spiral and scroll waves are robust solutions of excitable and oscillatory media in, respectively, two and three spatial dimensions.
Stability of these solutions is described by the spectra of the evolution operators obtained by linearizing the governing equations. 
Although the governing equations describing homogeneous and isotropic media respect Euclidean symmetry (on the \rgedit{$\mathbb{R}^2$} plane), the particular wave solutions do not. 
As a result, the \rgedit{respective} evolution operators are generically non-self-adjoint, so that their right (conventional) eigenfunctions are not mutually orthogonal and do not coincide with the left (adjoint) eigenfunctions.

The spectra of the evolution operators have a useful dynamical interpretation: the eigenvalues describe the temporal evolution of different perturbations, with positive (negative) real part corresponding to unstable (stable) modes, \cmedit{and} the eigenfunctions describe the spatial structure of the perturbation mode growing or decaying at the rate defined by the corresponding eigenvalue, while the adjoint eigenfunctions describe the sensitivity of different modes to \cmedit{perturbations in} the initial conditions.

Although non-trivial wave solutions do not respect the underlying Euclidean symmetry of the evolution equation, the symmetry manifests itself in the emergence of marginal modes (with zero eigenvalues) in the spectrum. In two dimensions the spectrum contains three marginal modes which correspond to the three continuous Euclidean symmetries: translation in the two directions spanning the plane and in-plane rotation\rgedit{\cite{Barkley94}}. The corresponding modes are known as Goldstone modes in Quantum Field Theory and Pattern Formation. The corresponding adjoint eigenfunctions have been termed response functions\rgedit{\cite{biktasheva1998}}. For stable spiral waves the Goldstone modes represent the dominant degrees of freedom and, in the presence of weak interactions, their evolution is naturally described in terms of the response functions. 

The earliest work illustrating the role of response functions in the context of reaction-diffusion systems is due to Keener~\cite{keener1988dynamics} who investigated the dynamics of scroll wave filaments. Scroll waves with a straight untwisted filament can be unstable even if the corresponding two-dimensional spiral wave solution is stable: the bend or twist of the filament leads to self-interaction which causes transverse motion of the filament that, for small curvature and torsion, can be described with the help of the response functions. 

The weak interaction picture also motivated investigations which aimed to describe the drift of stable spirals in two dimensions. Biktashev and Holden \cite{biktashev1993resonant} showed that experimental and numerical results describing the dynamics of a spiral wave in the presence of resonant forcing, perturbations of parameters, or interaction with the boundary can be understood using an empirical model containing three coupled ordinary differential equations (ODE) for the position of the wave core and the phase of the wave. In a subsequent paper the same authors~\cite{Biktashev:1995re} showed that the ODEs can be derived with the help of the three response functions.

The response functions themselves, however, were not computed for any reaction-diffusion system until much later. Biktasheva {\it et al.}~\cite{biktasheva1998} computed them for the spiral waves in the complex Ginzburg-Landau equation (CGLE), which describes the generic dynamics of a broad class of spatially extended systems close to the onset of the oscillatory instability, and Henry and Hakim \cite{Henry2000,Henry2002} computed response functions for scroll waves in the Barkley model of excitable media. In both instances the calculations were performed in a co-rotating reference frame which transforms the time-dependent solutions (relative equilibria) into steady states (absolute equilibria). This reduction based on the equivalence of the temporal evolution and rotation has been employed for computing the adjoint eigenfunctions in all subsequent studies.

With the help of the computed response functions it was possible to check that there is not just qualitative, but also quantitative agreement between numerical simulations and the ODE-based model for the resonant drift of spiral waves in CGLE~\cite{biktasheva1999resonant} in the Eckhaus-stable parameter regime. All three response functions were found to decay exponentially with the distance to the spiral core for the CGLE, in agreement with the analytical prediction. Exponential localization was later found even in the Eckhaus-unstable parameter regime~\cite{biktasheva2001}. 

Response functions can be used to describe the interaction of spiral waves not only with boundaries, but also with other spiral waves. Indeed, exponentially decaying interaction for spiral wave solutions of CGLE was predicted previously using the amplitude/phase equation formalism \cite{aranson1991interaction, pismen1991mobility, pismen1992interaction}. The analytical results obtained for CGLE, however, cannot be extended to strongly nonlinear waves in excitable media, so the response function formalism becomes the only tractable means of predicting the evolution of spiral waves in response to internal or external perturbations. 

Quite interestingly, the response functions were found to be exponentially localized also for the Barkley \cite{Henry2002}, FitzHugh-Nagumo \cite{BiHoBi06}, Oregonator \cite{biktasheva2015}, and Beeler-Reuter-Pumir \cite{Biktashev2011} models, suggesting that, as a rule, the spiral core acts as the organizing center for the wave, although there are rare counter-examples \cite{biktashev2005causodynamics} such as the Mornev model \cite{Mornev2003}. 

The exponential localization of the response functions enables quantitative description of the dynamics of the spiral core as a singular forced object. As Ref. \onlinecite{biktasheva2003wave} put it, ``spiral waves look like essentially nonlocalized objects but behave as effectively localized particles.'' As a result, despite the dissipative nature of excitable media, one finds a wave-particle duality that is more akin to that found in Hamiltonian systems. For example, Langham and Barkley \cite{LanBar13,LanBar14} used the response function formalism to show that core of a resonantly driven spiral in a bounded domain moves almost like a classical particle, although reflections from the ``walls'' are characterized by a strongly nonlinear relation between the incident and reflected angle.

When the spiral or scroll wave is unstable, its dynamics can not be described in terms of the marginal degrees of freedom (i.e., Goldstone modes and response functions). Instead, one also must consider the evolution of all the unstable modes. The only relevant study that we are aware of is due to Allexandre and Otani \cite{Otani2004} who considered the problem of feedback control of unstable spiral wave solutions of the Fenton-Karma~\cite{Fenton1998} model. In addition to the response functions, the eigenfunctions adjoint to all of the unstable modes were computed. The structure of the unstable adjoint eigenfunctions is especially important in the control problem as it allows significant optimization \cite{Garzon11,Garzon14}.

In all of the above studies only one type of spiral wave solution was considered -- those described by relative equilibria -- which can be reduced to a steady state in a rotating reference frame. However, most non-trivial applications involve more complex types of solutions. For instance, the simplest type of a spiral wave on a bounded domain of generic shape is described by a periodic solution. The simplest meandering spirals, even in \rgedit{$\mathbb{R}^2$}, are described by relative periodic solutions that are reducible to periodic solutions in a translating and rotating reference frame. The generalization of even the most basic results obtained for relative equilibria to more complex types of solutions is far from straightforward. Furthermore, it is not understood what kind of effects are introduced by the intrinsic time-dependence of the shape of a spiral wave.

In this work we present a generic method for computing the Floquet spectra, including both the right and left eigenfunctions, of essentially time-dependent spiral wave solutions, such as absolute and relative periodic orbits, on domains of generic shape. 
Unlike methods based on symmetry reduction, the presented approach allows one to compute all the leading modes in the spectrum, not just the Goldstone modes and the corresponding response functions, for both stable and unstable solutions.  
Sect.~\ref{sec:mp} gives the mathematical preliminaries including the description of the mathematical model. 
Sect.~\ref{sec:r} describes the properties of the Floquet eigenfunctions and their adjoints for drifting spiral waves described by (generalized) relative periodic solutions. 
The localization of adjoint eigenfunctions and the interaction of spiral waves with physical boundaries are discussed in Sect.~\ref{sec:d}.
We end with some general conclusions and summary in Sect.~\ref{sec:c}.

\section{Mathematical preliminaries \label{sec:mp}}

Our focus here is on the unstable spiral wave solutions that describe features of complex cardiac arrhythmias, such as fibrillation. Cardiac tissue is an example of excitable media, which are commonly modeled using reaction-diffusion equations,
\begin{equation}\label{eq:RDE}
	\partial_t\vec{u} = \cmedit{\nabla\cdot\sigma\nabla}\vec{u} + \vec{f}(\vec{u}).
\end{equation}
The interpretation of the dynamical variables $\vec{u}=[u_{1}\dots u_{l}](t,\vec{x})$ depends on the details of the system under consideration. In monodomain models of cardiac tissue these represent the transmembrane voltage, ionic fluxes, etc. The diffusion tensor $\sigma$ describes the intercellular couplings and the nonlinear function $\vec{f}(\cdot)$ describes the cellular dynamics. We use a modified form of the Karma model~\cite{Karma1993,karma94} \rgedit{with $l=2$ and:
\begin{align}
f_1&=(u^* - u_2^{M})\{1 - \tanh(u_1-3)\}u_1^{2}/2 - u_1 \nonumber\\
f_2&=\epsilon\left\{\beta \Theta_s(u_1-1) + \Theta_s(u_2-1)(u_2-1) - u_2\right\},
\label{eq:modKarma}
\end{align}
where $u_1$ is the non-dimensional transmembrane potential and $u_2$ is a gating variable.}
This system of equations was formulated as a minimal model of cardiac excitation dynamics exhibiting alternans -- \cmedit{the instability that is believed to be responsible for initiating and sustaining fibrillation\cite{Pastore1999,Jalife2000,ten2006alternans}.} 
\cmedit{Alternans manifests as a period-doubling bifurcation in isolated cells\cite{Nolasco1968} and a Hopf bifurcation in tissue~\cite{frame1988oscillations}}\rgedit{, where the temporal period and wavelength of the perturbation are nearly double those of the base excitation wave.}
The parameters of the model are as in Ref.~\onlinecite{Marcotte2015}, with the exception of the width of the switching function, which we set to $s=1.2571$ throughout.

\rgedit{For $\sigma=D=\mathrm{const}$, equation~\eqref{eq:RDE} with $\vec{x}\in\mathbb{R}^2$ and $t\in\mathbb{R}$} is equivariant with respect to continuous group actions from \rgedit{$E(1) \times E(2)$, consisting of temporal shifts $t\rightarrow t'=t+\tau$} and continuous translation and rotation transformations of the spatial coordinates $\vec{x} \rightarrow \vec{x}' = R_\phi\vec{x} + \vec{h}$, where $R_\phi=\exp(\phi\partial_\theta)$ is the operator of rotation by angle $\phi$ and $\partial_\theta$ is the generator of rotations. 
A consequence of this equivariance is that each solution lies on a  manifold of equivalent solutions, generated by the action of symmetry transformations on the state\cmedit{\cite{Barkley94}}. 
In particular, the Goldstone Modes correspond to the generators of  translational ($\partial_{\vec{x}}\vec{u}$) and rotational ($\partial_{\theta}\vec{u}$) symmetries in space and forward-translation symmetry in time ($\partial_t\vec{u}$) that span the tangent space of the group manifold\cmedit{\cite{BiHoNi96}}.

The presence of these symmetries, in turn, allows the special types of traveling wave solutions (i.e., plane and spiral waves) described by relative equilibria. 
In particular, the rigidly-rotating spiral waves which satisfy 
\begin{equation}\label{eq:req}
	\partial_t\vec{u}=\omega\partial_{\theta}\vec{u}
\end{equation}
can be found even on bounded domains as long as their shape is consistent with the rotational symmetry (e.g., circular domains)\cmedit{\cite{barkley1992}}. 
The equivalence between rotation and time-translation explicit in \eqref{eq:req} is the special property that enabled computation of the marginal modes by transforming the equation into a reference frame co-rotating with angular velocity $\omega$, which transforms a relative equilibrium into an absolute equilibrium. This equivalence is also the reason there are three, rather than four, independent Goldstone modes for relative equilibria.

Although circular domains are \cmedit{computationally and mathematically convenient}, they are \cmedit{physiologically irrelevant}. 
In particular, cardiac tissue is not circular, and in more complex regimes multi-spiral states partition the domain into tiles, each of which supports a single spiral \cite{ByMaGr14}, that never, even transiently, acquire a circular shape. 
Once the boundary loses circular symmetry, spiral wave solutions cannot be described by relative equilibria. 
Instead, one finds solutions such as pinned spirals (absolute periodic orbits) 
\begin{align}\label{eq:po}
\mathcal{U}_T\vec{u}(0,\vec{x})=\vec{u}(0,\vec{x})
\end{align}
or drifting spirals
\begin{align}\label{eq:rpo}
\mathcal{U}_T\vec{u}(0,\vec{x})=\mathcal{T}_{\vec{h}}\vec{u}(0,\vec{x})=\vec{u}(0,\vec{x}+\vec{h}),
\end{align}
where \rgedit{$\mathcal{U}_T$ is the time-$T$ evolution operator, and} $\mathcal{T}_{\vec{h}}=\exp(\vec{h}\cdot\nabla)$ is the translation operator. 
\cmedit{Relative periodic orbits on \rgedit{$\mathbb{R}^2$}  satisfy \eqref{eq:rpo} exactly and become periodic solutions in a reference frame moving with velocity $\vec{c}=\vec{h}/T$.
On bounded domains, these solutions are elevated to generalized relative periodic orbits which are {\it numerically} exact (i.e., exact to the level of numerical accuracy) only in the interior of the domain, outside of a narrow boundary layer \cite{Marcotte2015}. } 
\rgedit{On bounded domains, the time-dependence of spiral wave solutions} is essential and cannot be removed by any symmetry transformation, requiring a different approach for computing the spectrum.

In this work we will assume that the \rgedit{spatial} domain $\Omega$ is finite and has a square shape, and impose the \rgedit{typical for cardiac models ``no-flux'' boundary conditions
\begin{equation}
	\vec{n}\cdot \vec{j} = \vec{0},
	\label{eq:BC}
\end{equation}
where $\vec{j}=\sigma\nabla\vec{u}$ is the flux and}
${\bf n}$ is the outward-oriented normal to the boundary $\partial\Omega$ of the domain. 
This choice represents a generic situation where the boundary conditions explicitly break the translational and rotational symmetry. 
The square shape is also \rgedit{representative of} the tiling decomposition \rgedit{of multispiral solutions}, where each tile is bounded by a small number of intersecting smooth curves meeting at finite angles\cite{ByMaGr14}.

The evolution equation \eqref{eq:RDE} along with boundary conditions~\eqref{eq:BC} defines a solution 
\begin{equation}
	\vec{u}(t,\vec{x}) = \mathcal{U}_t\vec{u}(0,\vec{x})
\end{equation}
to the initial value problem, where $\mathcal{U}_t$ is the \cmedit{time-$t$} evolution operator of the nonlinear reaction-diffusion system \eqref{eq:RDE}. The temporal evolution of small perturbations $\vec{v}(t,\vec{x})$ about this solution is described by the linearization of \eqref{eq:RDE}, with the boundary conditions $\vec{n}\cdot\nabla\vec{v}(t,\vec{x}) = \vec{0}$ inherited from \eqref{eq:BC}. The resulting (forward) evolution equation in the tangent space is
\begin{align} \label{eq:right}
	\partial_t\vec{v}(t,\vec{x}) = L[\vec{u}(t,\vec{x})]\vec{v}(t,\vec{x}),
\end{align}
where $L[\vec{u}] = D{\nabla^2} + \vec{f}'(\vec{u})$ is the instantaneous Jacobian operator and $\vec{f}'(\vec{u}) = \partial \vec{f}/\partial \vec{u}$ describes the variation of the cellular dynamics with respect to the state variables.

The propagator \cmedit{$\mathcal{V}_t$} maps an initial infinitesimal perturbation $\vec{v}(0,\vec{x})$ in the tangent space to a later time, \cmedit{$\vec{v}(t,\vec{x}) = \mathcal{V}_t\vec{v}(0,\vec{x})$}, where
\begin{equation}\label{eq:tangentMap}
	\cmedit{\mathcal{V}_t = \exp{\left\{\int_0^t\dd t'\,L[\vec{u}(t',\vec{x})]\right\}}.}
\end{equation}
Similarly, the action of the adjoint propagator \cmedit{$\mathcal{V}_t^{\dagger}$} maps an infinitesimal \rgedit{perturbation $\vec{w}(t,\vec{x})$ in the adjoint space to an earlier time, $\vec{w}(0,\vec{x}) = \mathcal{V}_t^{\dagger}\vec{w}(t,\vec{x})$}, where
\begin{equation}
	\cmedit{\mathcal{V}_t^{\dagger} = \exp{\left\{\int_0^t \dd t'\,L^\dagger[\vec{u}(t-t',\vec{x})] \right\}},}
\end{equation}
and $L^\dagger[\vec{u}(t,\vec{x})]$ is the adjoint of the instantaneous Jacobian operator \eqref{eq:right}. It is easy to check that \cmedit{$\mathcal{V}_t^{\dagger}$ defines a time-reversed flow in the adjoint space
\begin{align} \label{eq:left}
	-\partial_t\vec{w}(t,\vec{x}) = L^\dagger[\vec{u}(t,\vec{x})]\vec{w}(t,\vec{x}).
\end{align}}

\rgedit{For $T$-periodic solutions,} the eigenfunctions and eigenvalues of $\mathcal{V}_T$ satisfy
\begin{align}\label{eq:floquet}
\vec{v}^i(T,\vec{x}) =\mathcal{V}_T\vec{v}^i(0,\vec{x})= \Lambda_{i}\vec{v}^i(0,\vec{x}).
\end{align}
Correspondingly,
\begin{align}
\vec{w}^i(0,\vec{x})=\mathcal{V}_T^{\dagger}\vec{w}^i(T,\vec{x}) = \Lambda_{i}^{*}\vec{w}^i(T,\vec{x}), 
\end{align}
where the asterisk denotes the complex conjugate. In other words $\vec{w}^i(T,\vec{x})$ and $\vec{v}^i(0,\vec{x})$ represent the left and right eigenfunctions of $\mathcal{V}_T$ corresponding to the eigenvalue $\Lambda_i$. 
Since $\mathcal{V}_T$ is generally non-self-adjoint, $\vec{w}^i(T,\vec{x})\neq \vec{v}^i(0,\vec{x})$, however the two sets of eigenfunctions are mutually orthogonal. The orthogonality relation can be written in a more useful and general form 
\begin{align}\label{eq:biorth}
	\langle \vec{w}^j(t) | \vec{v}^i(t) \rangle = \int_{\Omega}\dd^{2}\vec{x}\, 
	[\vec{w}^j(t,\vec{x})]^\dagger\vec{v}^i(t,\vec{x})
	 = \delta_{ij}
\end{align}
for $0\leq t\leq T$, where $\vec{v}^i(t,\vec{x})$ is defined as the solution of \eqref{eq:right} that coincides with the right eigenfunction $\vec{v}^i(0,\vec{x})$ at $t=0$ and $\vec{w}^j(t,\vec{x})$ as the solution of \eqref{eq:left} that coincides with the left eigenfunction $\vec{w}^j(T,\vec{x})$ at $t=T$. For a time-periodic solution $\vec{u}(t,\vec{x})$ with period $T$, $\vec{v}^i(t,\vec{x})$ and $\vec{w}^j(t,\vec{x})$ correspond to the Floquet modes.

Although the vast majority of studies of infinite-dimensional systems focus exclusively on the right (conventional) eigenfunctions, the importance of the left (adjoint) eigenfunctions is hard to overestimate. Indeed, an accurate estimate of the evolution operator $\mathcal{V}_T$ is crucial for a number of applications such as computing unstable solutions \cite{Marcotte2015}, feedback control \cite{Otani2004,Garzon11,Garzon14}, and adjoint-based optimization. For infinite-dimensional systems, explicit evaluation of $\mathcal{V}_T$ is prohibitively expensive (or simply impossible). Such estimates are often constructed by using Arnoldi iterations\rgedit{\cite{Arnoldi1951}} which generate approximations of the leading right eigenfunctions, but generally do not approximate the left eigenfunctions. An optimal estimate based on the truncated spectral decomposition \begin{align}
	\mathcal{V}_T\approx\sum_{i=1}^N|\vec{v}^i\rangle\Lambda_i\langle \vec{w}^i|
\end{align}
obviously requires both sets of eigenfunctions. In comparison, an Arnoldi-based approximation
\begin{align}
	\mathcal{V}_T\approx\sum_{i=1}^M|\vec{v}^i\rangle\Lambda_i\langle \bar{\vec{w}}^i|,
\end{align}
\rgedit{e.g., used in GMRES methods\cite{saad1986gmres},}
uses synthetic adjoints $\bar{\vec{w}}^i$ which satisfy \eqref{eq:biorth} only in the $M$-dimensional Krylov subspace spanned by $\{\vec{v}^1,\cdots,\vec{v}^M\}$ (i.e., $\langle \bar{\vec{w}}^{j} | \vec{v}^{i} \rangle\neq \delta_{ij}$ if either $i$ or $j>M$) and requires $M\gg N$ to achieve a similar level of accuracy. 
Similarly, the adjoints are required to compute the coordinates $a_i$ of disturbances in the tangent space:
\begin{equation}\label{eq:coord}
	\delta\vec{u} = \sum_{i} a_{i}\vec{v}^i, \qquad a_{i} = \langle \vec{w}^i | \delta\vec{u} \rangle.
\end{equation}

\section{Numerical Results \label{sec:r}}

\begin{figure}[t]
	\includegraphics[width=\columnwidth]{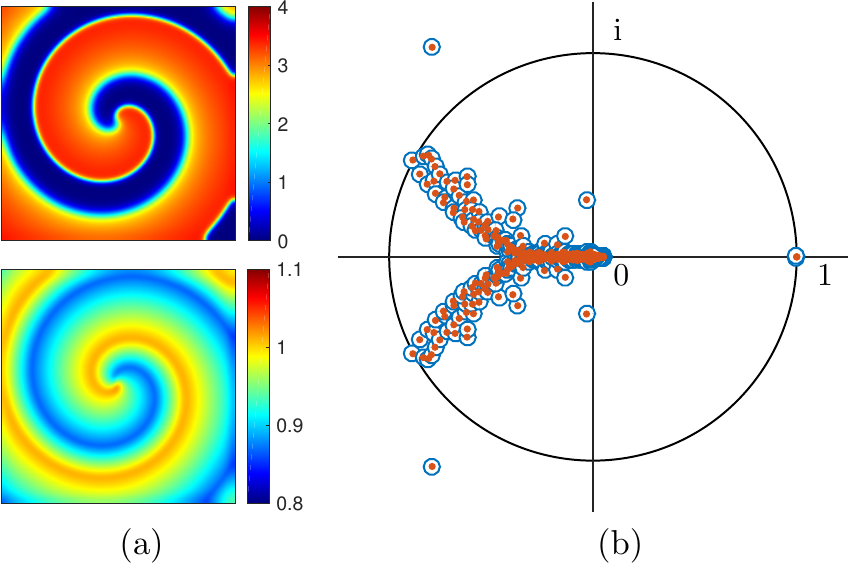}
	\caption{
	(a) Snapshot $\vec{u}(0,\vec{x})$ of the unstable generalized relative periodic solution with period $T=54.74$ and wavelength $\lambda=78$ (the first and second component are shown above, and below, respectively).
	(b) Floquet spectrum of the solution from the right (\cmedit{blue circles}) and left (\cmedit{red dots}) eigenfunction calculations.
	\label{fig:spectrum}}
\end{figure}

In a previous study \cite{Marcotte2015} we have introduced a numerical method \rgedit{which employs a Newton-Krylov solver} for computing both the \rgedit{unstable} generalized relative periodic orbits of \eqref{eq:RDE} and their right eigenfunctions on bounded domains of arbitrary shape that break all the continuous symmetries of the underlying evolution equations. 
A snapshot of a single-spiral solution described by a generalized relative periodic orbit is presented in Fig.~\ref{fig:spectrum}(a).
This spiral wave solution has wavelength $\lambda = 78$ ($2.04$~cm) 
and period $T = 54.74$ ($136.9$~ms) 
on a domain of size $L=192$ ($5.03$~cm). 
\rgedit{The unit conversion uses the typical size of a cardiac cell ($262$ $\mu$m) for nondimensionalization.}
When the tip $\vec{x}_{o}$ of the spiral (defined by $\partial_t\vec{u}(0,\vec{x}_{o}) = \vec{0}$) is placed within $10$~$\mu$m of the center of the domain, it drifts by just $|\vec{h}| = O(10^{-11}\lambda)$ over the course of the rotation, reflecting the discrete rotational symmetry of the problem associated with this particular initial condition.
Hence, although it is a member of a class of relative periodic orbits, this particular solution is, to numerical precision, simply a periodic orbit.
For other choices of $\vec{x}_o$, we found that $|\vec{h}|\propto e^{-\zeta/\ell_{c}}$, where $\zeta$ is the distance from $\vec{x}_o$ to the nearest boundary and $\ell_{c}$ is a numerically determined critical length scale.

This study extends the results of Ref. \onlinecite{Marcotte2015} by using a method (detailed in the Appendix) that does not rely on a transformation to the co-moving frame, thus allowing computation of adjoint eigenfunctions for spiral wave solutions with arbitrary symmetry properties and temporal dependence. 
The spectrum of Floquet multipliers $\Lambda_i$ for the spiral wave solution is shown in Fig.~\ref{fig:spectrum}(b). The Floquet multipliers correspond to the eigenvalues of $\mathcal{V}_T$; the corresponding Floquet exponents $\sigma_i$ can be computed using the relation $\Lambda_i=e^{\sigma_iT}$.
The spectrum is seen to include both a discrete and a continuous part.
There are four unstable eigenvalues 
\rgedit{(two from the discrete and two from the continuous part)} as well as a triply-degenerate marginal eigenvalue ($\Lambda = 1$) associated with infinitesimal spatial and temporal translations, just like for relative equilibria\rgedit{\cite{Barkley94}}. 
The two Goldstone modes associated with infinitesimal spatial translations persist on bounded domains \cite{Barkley94,Marcotte2015} due to the {\it local} translational invariance of \eqref{eq:RDE} even though the {\it global} translational symmetry is broken by the boundary conditions \eqref{eq:BC}. 

\begin{figure}[t]
	\includegraphics[width=\columnwidth]{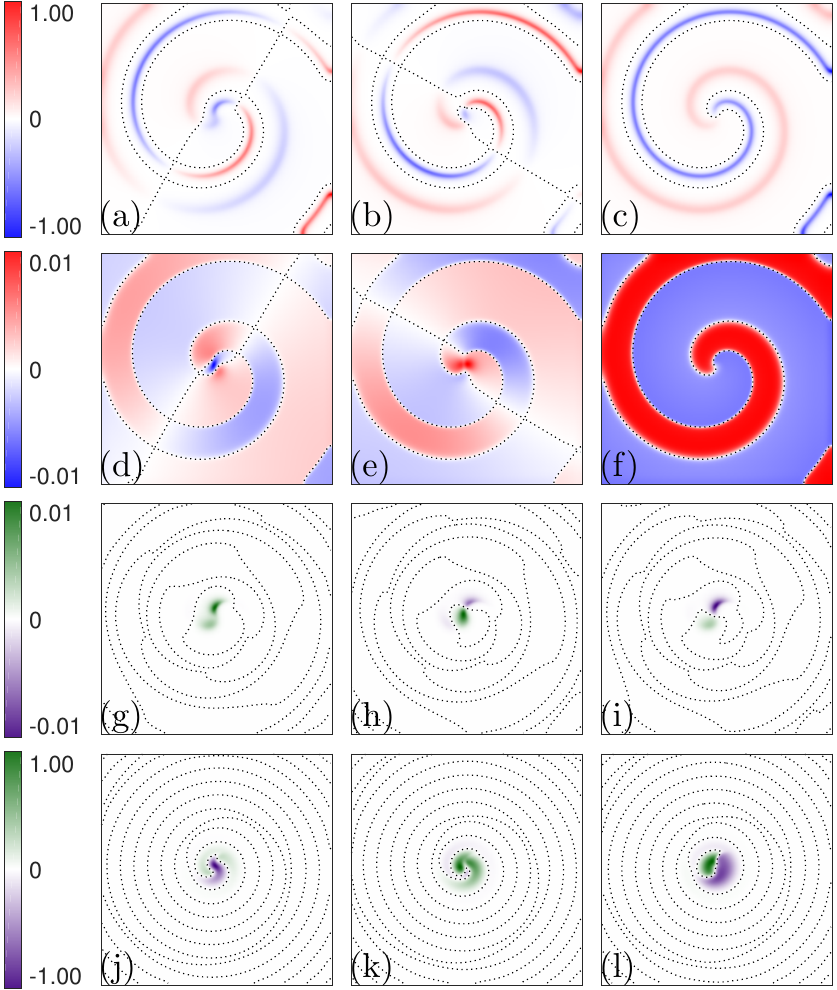}
	\caption{Goldstone modes (a-f) and response functions (g-l). 
	The left two columns show the modes associated with spatial translation and the right column -- the modes associated with temporal translations. 
	The first (a-c) and second (d-f) rows show, respectively, the first and second components $v^i_1$ and $v^i_2$ of the Goldstone modes. 
	The third (h-i) and fourth (j-l) rows show, respectively, the first and second components $w^i_1$ and $w^i_2$ of the response functions.
	The dotted curves denote nodal lines. 
	Here and below the snapshots of eigenfunctions are shown at the same time instant ($t=0$) as the spiral wave solution in Fig.~\ref{fig:spectrum}.
\label{fig:GM+RF}}
\end{figure}

We have computed the leading $130$ left and right \rgedit{eigenmodes with high accuracy (see Appendix).} 
The movies of $\vec{v}^i(t,\vec{x})$ and $\vec{w}^j(t,\vec{x})$ for several dominant modes are included in the supplemental material (Sect.~\ref{sec:sm}).
\rgedit{As Fig.~\ref{fig:spectrum}(b) shows, the} eigenvalues associated with the left eigenfunctions are just as accurate as those associated with the right eigenfunctions.
In particular, the eigenvalues associated with the marginal modes deviate from unity less than $O(10^{-7})$ for both the Goldstone modes and the response functions. 

The Goldstone modes associated with translational symmetries are shown in Fig. \ref{fig:GM+RF}(a-b) and (d-e).
These correspond to the spatial derivatives of the initial condition $\vec{n}\cdot\nabla\vec{u}$ along two orthogonal directions $\vec{n}$.
Figures \ref{fig:GM+RF}(c) and (f) shows the Goldstone mode associated with temporal translation which corresponds to the temporal derivative $\partial_t\vec{u}$.
Figures~\ref{fig:GM+RF}(g-l) show the corresponding response functions.
The orthogonality condition \eqref{eq:biorth} does not completely fix the normalization of the two sets of eigenfunctions. 
\cmedit{As the eigenfunctions represent solutions to a linear problem (and are thus scale-independent), we are free to \rgedit{choose} the absolute scale of each set, provided \eqref{eq:biorth} is satisfied.}
Hence, we added an additional constraint $\langle\vec{v}^i|\vec{v}^i\rangle = \langle \vec{w}^i|\vec{w}^i\rangle$, so that the \cmedit{scales of the dominant components of $\vec{v}^i$ and $\vec{w}^i$} are comparable. 

\begin{figure}[t]
	\includegraphics[width=\columnwidth]{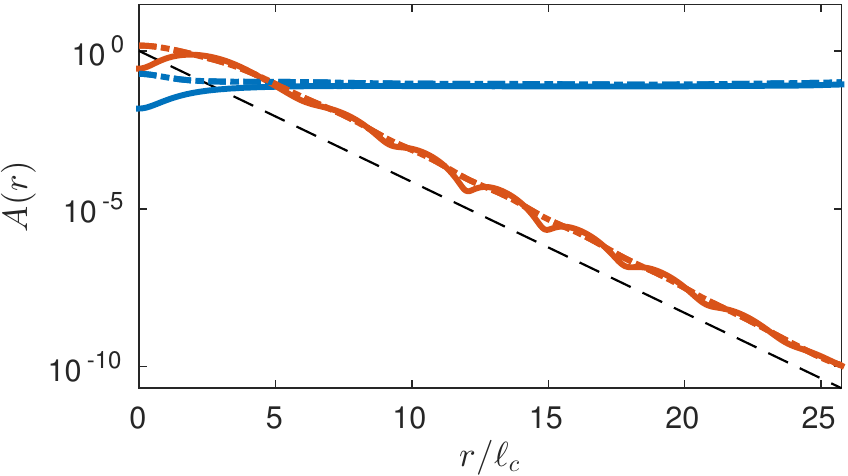}
	\caption{Amplitude of the left (\cmedit{red}) and right (\cmedit{blue}) eigenfunctions corresponding to spatial translations (solid), and temporal translations (dashed). 
	Numerically determined scaling for the response functions $\ell_{c}^{\mathrm{num}} = \lambda/20$ (black, dashed).
	\label{fig:RF+GM-A}}
\end{figure}

As Fig. \ref{fig:GM+RF} illustrates, for the Karma model the second component (gating variable) of the right eigenfunctions is very small compared to the first component (voltage variable), $\|v^i_2\|=O(10^{-2}\|v^i_1\|)$, while for the left eigenfunctions the opposite is true, $\|w^i_1\|=O(10^{-2}\|w^i_2\|)$. 
This disparity is due to the difference in the time scales of cellular kinetics described by the nonlinear functions ${\bf f}=(\tilde{f}_1,\epsilon\tilde{f}_2)$, where $\epsilon=0.01$ is a fixed parameter and both functions  $\tilde{f}_1$ and $\tilde{f}_2$ and their partial derivatives are all $O(1)$. For instance, if we rescale the two components of the right eigenfunction as $\vec{v}^i=(\tilde{v}^i_1,\epsilon\tilde{v}^i_2)$, then the evolution equation \eqref{eq:right} can be rewritten as 
\begin{align}\label{eq:right_eps}
\partial_t\tilde{v}^i_1&=D_{11}\nabla^2\tilde{v}^i_1
+\frac{\partial\tilde{f}_1}{\partial{u}_1}\tilde{v}^i_1
+\epsilon\frac{\partial\tilde{f}_1}{\partial{u}_2}\tilde{v}^i_2,\nonumber\\
\partial_t\tilde{v}^i_2&=D_{22}\nabla^2\tilde{v}^i_2
+\frac{\partial\tilde{f}_2}{\partial{u}_1}\tilde{v}^i_1
+\epsilon\frac{\partial\tilde{f}_2}{\partial{u}_2}\tilde{v}^i_2,
\end{align}
where $\partial \tilde{f}_i/\partial u_j=O(1)$ for all $i$ and $j$.
Similarly, rescaling the components of the left eigenfunctions $\vec{w}=(\epsilon\tilde{w}^i_1,\tilde{w}^i_2)$, we can rewrite the evolution equation \eqref{eq:left} as 
\begin{align}\label{eq:left_eps}
-\partial_t\tilde{w}^i_1&=D_{11}\nabla^2\tilde{w}^i_1
+\frac{\partial\tilde{f}_1}{\partial{u}_1}\tilde{w}^i_1
+\frac{\partial\tilde{f}_2}{\partial{u}_1}\tilde{w}^i_2,\nonumber\\
-\partial_t\tilde{w}^i_2&=D_{22}\nabla^2\tilde{w}^i_2
+\epsilon\frac{\partial\tilde{f}_1}{\partial{u}_2}\tilde{w}^i_1
+\epsilon\frac{\partial\tilde{f}_2}{\partial{u}_2}\tilde{w}^i_2.
\end{align}
Equations \eqref{eq:right_eps} and \eqref{eq:left_eps} have solutions both components of which are of the same order of magnitude, so we should indeed expect $\|v^i_2\|=O(\epsilon)\|v^i_1\|$ and $\|w^i_1\|=O(\epsilon)\|w^i_2\|$ for all eigenfunctions. Hence, in the remainder of the paper we will focus mostly on their dominant (unscaled) components $v^i_1$ and $w^i_2$.

The most salient feature of the response functions in the Karma model is that they are very strongly localized near the core of the spiral, just like in most other excitable systems and the CGLE. To quantify this spatial localization we defined the amplitude
\cite{BBBBF09}
\begin{equation}
	A(r) =\left(\frac{1}{2\pi}\int_0^{2\pi} |\vec{w}(r,\theta)|^{2}\,\dd\theta\right)^{1/2},
\end{equation}
where $r$ is the distance from $\vec{x}_{o}$. 
The amplitude can be defined in a similar manner for all eigenfunctions, both left and right.
As Fig.~\ref{fig:RF+GM-A} shows, the response functions decay exponentially with  $r$, while the amplitude of the Goldstone Modes, as expected, remains a constant outside of the core region. For comparison, the dashed line shows the exponential decay $A\propto e^{-r/\ell_{c}}$ predicted numerically \cite{Marcotte2015} based on the scaling results for the spatial drift ${\bf h}$ and the period $T$ of the spiral wave, where $\ell_{c} \approx \lambda/20$ for the value of $s$ used here. The spatial decay rate determined directly from the response functions gives a very close value
$\ell_{c} = 0.0478\lambda \approx \lambda/21$, which confirms the conjecture \cite{Marcotte2015} that the scaling of ${\bf h}$ and $T$ is indeed controlled by the spatial structure of the response functions corresponding, respectively, to the spatial and temporal Goldstone modes.

\begin{figure}[t]
	\includegraphics[width=\columnwidth]{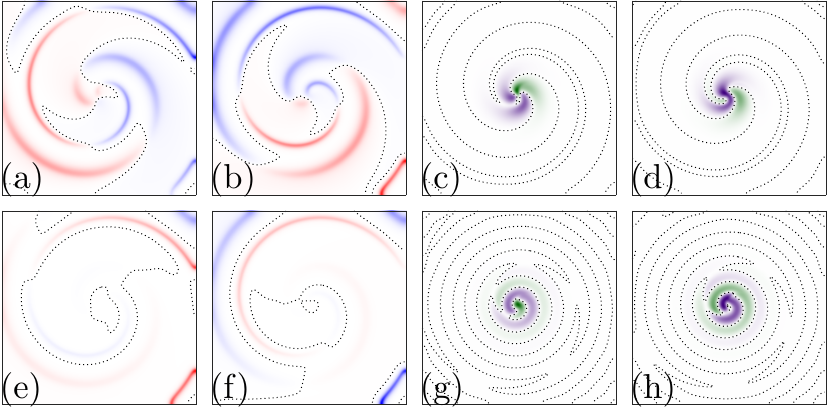}
	\caption{Snapshots of the unstable right (a-b) and corresponding left (c-d) eigenfunctions from the discrete spectrum with Floquet multiplier $\Lambda_\pm = -0.7893\pm1.0286\ii$.
	Snapshots of the stable right (e-f) and corresponding left (g-h) eigenfunctions from the discrete spectrum with Floquet multiplier $\Lambda_\pm = -0.0315\pm0.2803\ii$.
	\label{fig:DS}}
\end{figure}

The response functions do not decay with $r$ monotonically. Instead, the dominant component $w_2$ displays pronounced oscillations, with roughly the same distance ($\Delta r \approx 3.04\ell_{c}$) between the nodal lines for all three response functions. 
For the temporal response function the nodal lines form closed loops in the plane. 
In contrast, the nodal lines of the translational response functions form spirals (they are not closed curves). Correspondingly, the angular averaging destroys the underlying oscillation of the amplitude $A(r)$ in the latter case, while in the former case the amplitude clearly shows the modulation superimposed on top of the exponential profile. 

\begin{figure}[t]
	\includegraphics[width=\columnwidth]{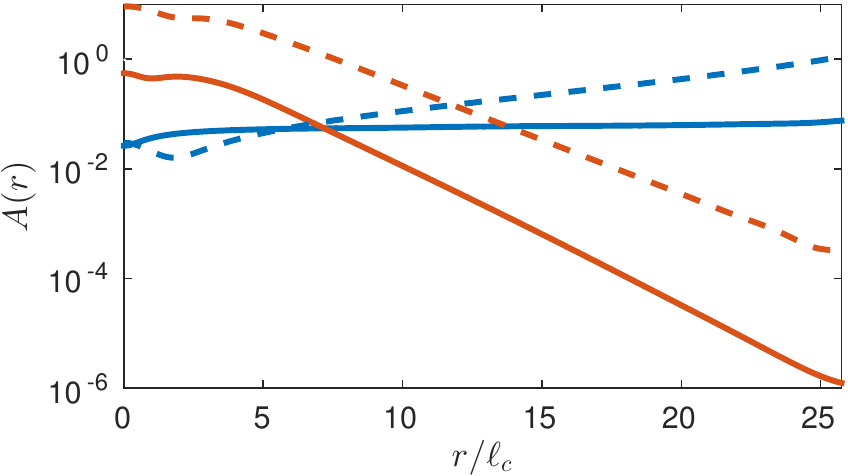}
	\caption{The amplitude of the \cmedit{left (red) and right (blue)} unstable (solid) and stable (dashed)  eigenfunctions shown in Figure~\ref{fig:DS}(a-b) \& (e-f) and Figure~\ref{fig:DS} (c-d) \& (g-h), respectively.
	\label{fig:DS-A}}
\end{figure}

The method introduced here allowed us to compute not only the marginal eigenfunctions, but an entire spectrum of leading modes. 
In particular, two pairs of complex conjugate modes -- the most unstable pair and a stable pair from the discrete part of the spectrum -- are shown in Figure~\ref{fig:DS} and  the corresponding angle-averaged amplitudes are plotted in Fig.~\ref{fig:DS-A}. 
Again we find that the adjoint eigenfunctions are strongly localized in the core region. Their amplitude {\it decreases} exponentially, $A(r)\propto e^{-r/\ell_{-}}$, just as it does for the response functions, albeit more slowly: the corresponding length scales are $\ell_{-}=1.72\ell_{c}$ and $\ell_{-}=2.20\ell_{c}$ for the unstable and stable modes, respectively. 
The right eigenfunctions show the opposite trend, their amplitude {\it increases} exponentially, $A(r)\propto e^{r/\ell_{+}}$. 
In  particular, the stable mode is localized near the boundary, with amplitude growing on the length scale $\ell_{+}=7.04\ell_{c}$.

The length scale $\ell_{+}\approx 5\lambda$ for the unstable modes is extremely large, so they appear to be evenly distributed throughout the entire domain (cf. Fig. \ref{fig:DS}(a-b)). 
The shape of the modes clearly indicates that they describe the alternans instability which is characterized by the variation in the width of the excitation wave. 
This particular pair of modes describes {\it discordant} alternans: as Fig. \ref{fig:DS} (a) illustrates, at $t=0$ the thickness \cmedit{of the excitation wave} increases in some regions (positive values of $\mathrm{Re}(v_1)$) and decreases in others (negative values of $\mathrm{Re}(v_1)$), which corresponds to an increase (respectively, descrease) in the action potential duration (APD). 
The corresponding Floquet multipliers are complex ($\mathrm{arg}(\Lambda) \approx \pm 2\pi/3$), rather than real and negative, as would be the case for a period-doubling bifurcation. 

This is, however, not the only unstable mode.
There are both {\it absolutely} unstable modes (characterized by $|\Lambda|>1$ or $\mathrm{Re}(\sigma)>0$) and {\it convectively} unstable modes (for which $\mathrm{Re}(\sigma)<0$) \rgedit{in the continuous part of the spectrum}. A right eigenfunction with amplitude $A(r)>C e^{r/\ell_{+}}$, where $C$ and $\ell_{+}$ are some positive constants, describes a convectively unstable mode, if $-c/\ell_{+}<\mathrm{Re}(\sigma)<0$, where $c=\lambda/T$ is the asymptotic conductive velocity of the spiral wave. Equivalently, the temporal decay $\mathrm{Re}(\sigma)T$ must overwhelm the spatial growth $\lambda/\ell_{+}$ in time $T$ and length $\lambda$ for the mode to be convectively stable.
\begin{figure}[t]
	\includegraphics[width=\columnwidth]{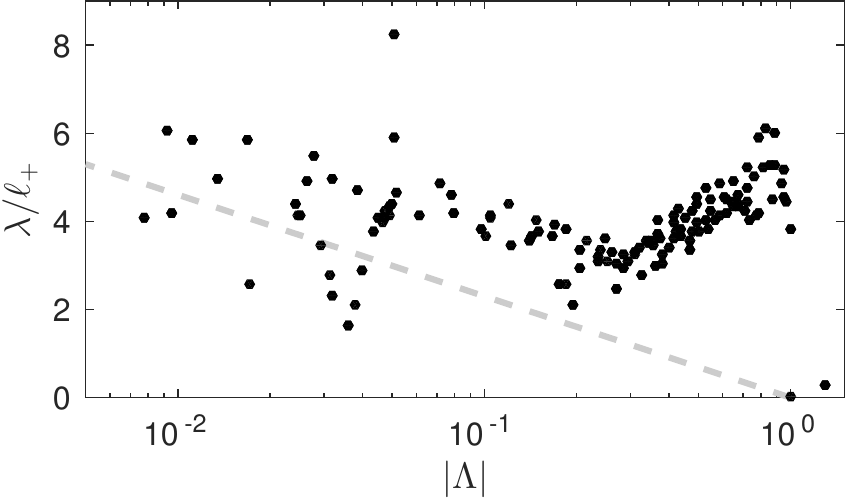}
	\caption{The spatial growth rates of the right eigenfunctions. Convectively unstable modes lie above the dashed line ($\lambda/\ell_{+}=-\mathrm{Re}(\sigma)T$) 	and to the left of $|\Lambda| = 1$.
		\label{fig:CI}}
\end{figure}
Convectively unstable modes would produce a noticeable distortion of the spiral wave on sufficiently large domains.
Unexpectedly, most of the absolutely stable modes computed for the solution shown in Fig.~\ref{fig:spectrum} grow with radial distance from the core so quickly that they are convectively unstable. As Fig.~\ref{fig:CI} illustrates, it is only a subset of the strongly contracting modes ($|\Lambda| < O(10^{-1})$) associated with relatively featureless eigenfunctions (i.e., for which $\ell_{+} = O(\lambda)$) which are both convectively and absolutely stable.
\rgedit{Convective instabilities and exponential growth of eigenfunctions for defect-modulated waves in reaction-diffusion systems have been investigated previously in one~\cite{Sandstede:2000ab} and two \cite{Sandstede:2000abs,Wheeler:2006co} spatial dimensions, and their role is reasonably well-understood. }

\begin{figure}[t]
	\includegraphics[width=\columnwidth]{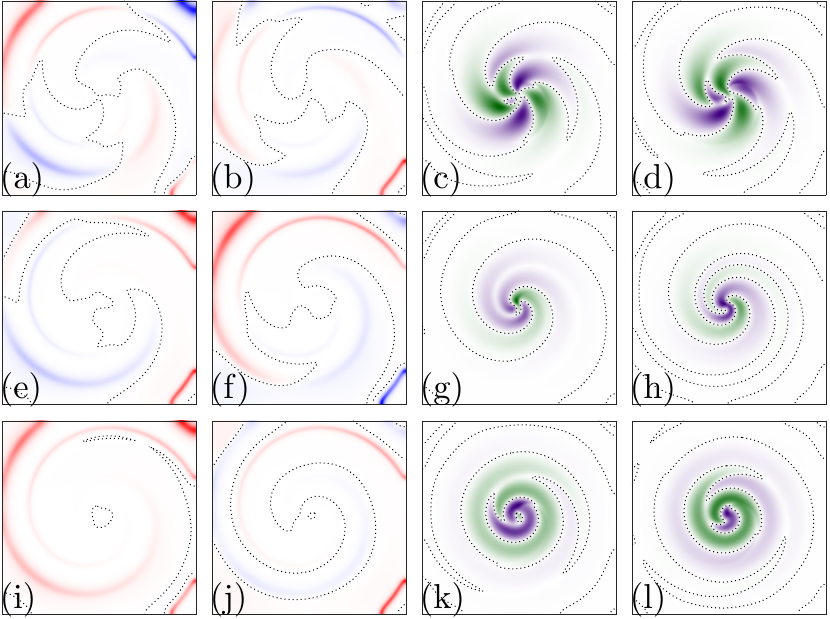}
	\caption{Snapshots of representative complex pairs of right eigenfunctions (a-b), $\Lambda = -0.8838 \pm 0.4753\ii$; 
	(e-f), $\Lambda = -0.8327\pm 0.4944\ii$; (i-j), $\Lambda = -0.8061\pm 0.4986\ii$) from the continuous spectrum near the unit circle, and their respective adjoint eigenfunctions (c-d), (g-h), and (k-l).
	\label{fig:CS}}
\end{figure}

\begin{figure}[t]
	\includegraphics[width=\columnwidth]{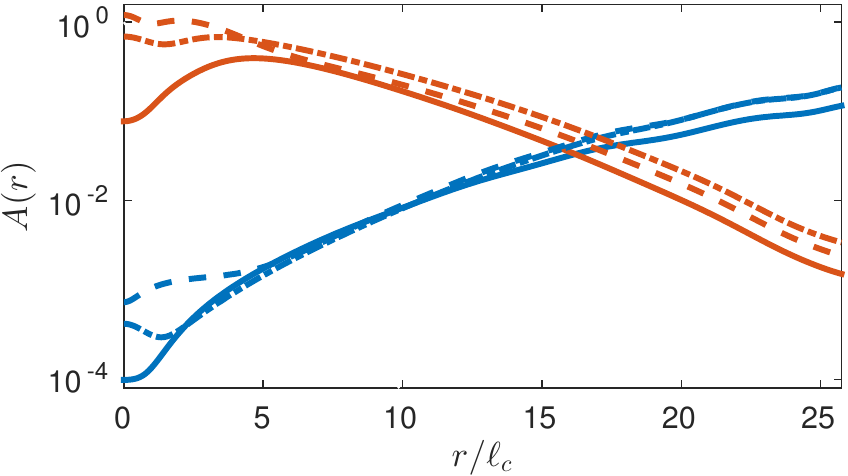}
	\caption{The \cmedit{radial} amplitudes of the left (\cmedit{red}) and right (\cmedit{blue}) eigenfunctions for the modes shown in Figure~\ref{fig:CS}(a-d, solid), (e-h, dashed), (i-l, dotted).
	\label{fig:CS-A}}
\end{figure}

As Fig. \ref{fig:spectrum} shows, the continuous spectrum crosses the unit circle near $\Lambda = e^{\pm 5\pi i/6}$, indicating that on \rgedit{$\mathbb{R}^2$}, one would expect to find an infinite number of modes close to the boundary of absolute instability. 
For the relatively small domain size considered here ($L = 2.46\lambda$), only a single complex pair of modes from the continuous spectrum is absolutely unstable (cf. Fig. \ref{fig:CS} (a-d)). 
However, the continuous spectrum contains a large number of modes that are convectively unstable (two examples are shown in Fig. \ref{fig:CS} (e-l)).
Figures~\ref{fig:CS} and \ref{fig:CS-A} show the eigenfunctions of three leading modes from the continuous spectrum and their amplitude. 

The leading modes in the continuous spectrum \cmedit{exhibit spatial localization trends similar to those from the discrete spectrum.}
In particular, the left eigenfunctions are localized in the core region, while the right eigenfunctions are localized near the boundaries (cf. Fig.~\ref{fig:CS}). 
The mode amplitudes, however, are not given by pure exponentials, but rather a product of an exponential and a power, i.e., 
\begin{align}\label{eq:exp-pow}
A(r)\propto (r/\ell_{\pm})^\alpha e^{\pm r/\ell_{\pm}}
\end{align}
with $\alpha \approx 2$ and $\ell_{-} \approx 2.5\ell_{c}$ for the left modes, and $\ell_{+} \approx 16\ell_{c}$
for the right modes, with $\Lambda_\pm = -0.88 \pm 0.48\ii$, $-0.83 \pm 0.50\ii$, and $-0.81 \pm 0.50\ii$. The spatial structure of the right eigenfunctions suggests that these modes are also related to alternans, although the width variation is mixed with bending of the excitation wave \rgedit{(in some regions both the leading and the trailing shock are displaced in the same direction, rather than in the opposite directions, as would be the case for alternans)}. 

To sum up, for the domain size considered here, there are several alternans modes in the Karma model.
Classical alternans (cf. Fig. \ref{fig:DS} (a-b)), the dominant mode of instability, lacks spatial localization, while there is another \cmedit{unstable} alternans mode that is localized near the boundary (cf. Fig. \ref{fig:CS} (a-b)). The adjoints of these modes are all strongly localized near the core.
In comparison, in the three-variable Fenton-Karma model \cite{Fenton1998}, the alternans modes of an unstable spiral wave computed on a disk of comparable size \cite{Otani2004} are characterized by adjoint eigenfunctions that show almost no attenuation with $r$. 
Our results show that, unlike the Fenton-Karma model where the development of alternans appears to be sensitive to perturbations over the entire domain, in the Karma model the development of alternans is only sensitive to perturbations near the spiral core.

\begin{figure}[t]
	\includegraphics[width=\columnwidth]{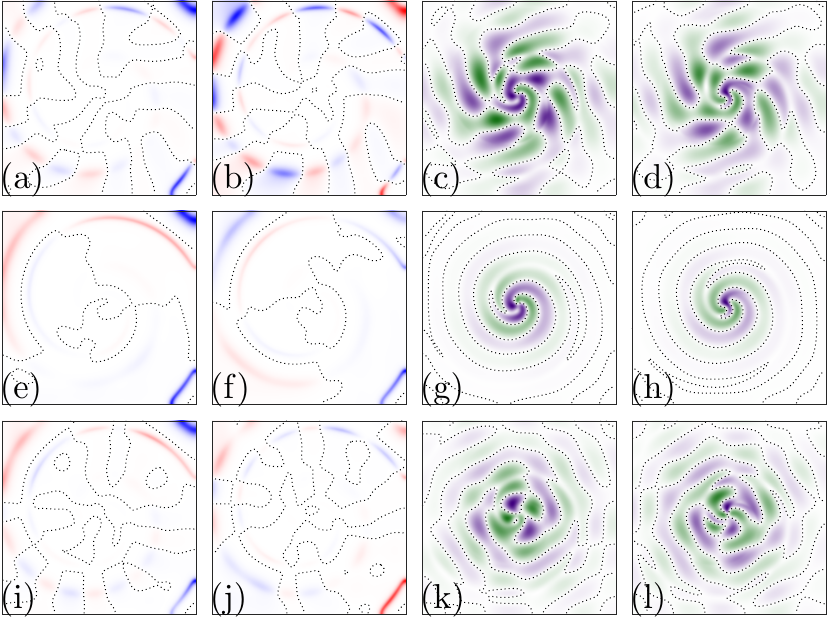}
	\caption{Strongly stable right (a,b,e,f,i,j) and left (c,d,g,h,k,l) eigenfunctions with multipliers $0.4 < |\Lambda| < 0.6$ (ordered by decreasing modulus).
	\label{fig:CS-S}}
\end{figure}

\begin{figure}[t]
	\includegraphics[width=\columnwidth]{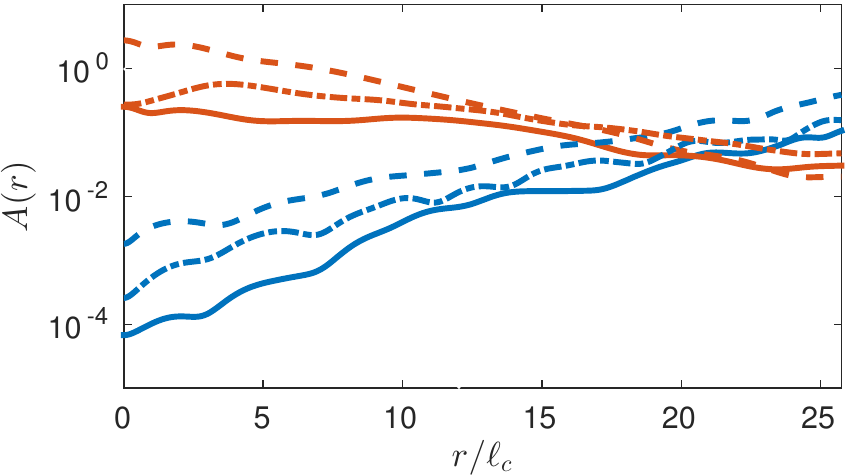}
	\caption{The \cmedit{radial} amplitudes \cmedit{for the right (blue) and left (red)} eigenfunctions shown in Fig. \ref{fig:CS-S} (a,d) solid, (b,e) dashed, (c,f) dash-dotted .
	\label{fig:CS-S-A}}
\end{figure}

In the conclusion of this section, we discuss the structure of some of the (absolutely) stable modes from the continuous part of the spectrum. The dominant components of the modes with Floquet multipliers $0.4 < |\Lambda| < 0.6$ are shown in Fig. \ref{fig:CS-S}. 
Similar to the unstable and weakly stable modes, the left eigenfunctions are found to be localized near the core of the spiral and the right eigenfunctions near the domain boundary, although the localization is weaker than for the modes with larger $|\Lambda|$. The amplitudes (cf. Fig. \ref{fig:CS-S-A}) are again found to increase/decrease exponentially (aside from some weak modulation) with $r$ on length scales $\ell_{\pm} = O(6\ell_{c})$.

\section{Discussion \label{sec:d}}

Although spatial localization of adjoint eigenfunctions appears to be an almost universal property of spiral wave solutions, it has only been understood for the response functions associated with a spiral wave solution in the CGLE \cite{biktasheva1998,biktasheva2001}. Extending these results for generic excitable systems has proved difficult due to the strong nonlinearity of the evolution equations. However, we can make progress in certain limits. Although the spiral wave solution investigated here formally corresponds to a relative periodic orbit, it is nearly indistinguishable from a rigidly rotating spiral wave (i.e., relative equilibrium) inside a circle of radius $L/2$. On an infinite domain, far from the origin, the Archimedian approximation applies
\begin{equation}\label{eq:Arc}
{\bf u}(r,\theta,t)\approx{\bf u}_0(\xi),
\end{equation}
where $\xi=r+l\theta-ct$, $l=\lambda/(2\pi)$, $c=\lambda/T$, and ${\bf u}_0(\xi)$ is periodic with period $\lambda$. The adjoints can then be written in the form $\tilde{\bf w}(r,\theta,t)=\bar{\bf w}(\xi) e^{im\theta} e^{\gamma t}$ and the second of the two equations in \eqref{eq:left_eps} becomes
\begin{align}\label{eq:w2_xi}
-\gamma\bar{w}^i_2&=D_{22}\left(\partial_\xi^2\bar{w}^i_2 +
r^{-1}\partial_\xi\bar{w}^i_2 -
m^2r^{-2}\bar{w}^i_2\right)\nonumber\\
&-c\partial_\xi\bar{w}^i_2+\epsilon\frac{\partial\tilde{f}_1}{\partial{u}_2}\bar{w}^i_1
+\epsilon\frac{\partial\tilde{f}_2}{\partial{u}_2}\bar{w}^i_2.
\end{align}
For $r \gg \lambda$, the curvature of the spiral wave can be ignored, and \eqref{eq:w2_xi} simplifies, yielding 
\begin{align}\label{eq:w2_xi_inf}
-\gamma\bar{w}^i_2&=D_{22}\partial_\xi^2\bar{w}^i_2-c\partial_\xi\bar{w}^i_2+\epsilon\frac{\partial\tilde{f}_1}{\partial{u}_2}\bar{w}^i_1
+\epsilon\frac{\partial\tilde{f}_2}{\partial{u}_2}\bar{w}^i_2.
\end{align}
Since the partial derivatives $\partial\tilde{f}_i/\partial{u}_j$ depend only on ${\bf u}_0$, they are periodic in $\xi$ and, according to the Floquet theorem, equation \eqref{eq:w2_xi_inf} has solutions in the form $\bar{w}^i_2(\xi)=\hat{w}^i_2(\xi)e^{k_i\xi}$, where $\hat{w}^i_2(\xi+\lambda)= \hat{w}^i_2(\xi)$. 
We therefore should expect the adjoints to grow or decay exponentially with $\xi$ or $r$ as long as $\mathrm{Re}(k_i)\ne 0$.
\rgedit{This result generalizes the prediction of exponential far-field dependence of left eigenfunctions  in one spatial dimension \cite{Sandstede:2000abs,sandstede2004defects}.}
\rgedit{Note that $\gamma=\sigma_i^*+ck_i$, where}
\begin{align}
\sigma_i=\frac{1}{T}\ln\Lambda_i=\frac{1}{T}\ln|\Lambda_i|+\frac{1}{T}\mathrm{arg}(\Lambda_i)
\end{align}
\rgedit{is the corresponding Floquet exponent.}

We can make further progress in various special cases. For instance, when $\epsilon\ll |\sigma_i|$ (e.g., for strongly stable modes), we have $\hat{w}^i_2=\mathrm{const}$ and $\sigma_i^*=-D_{22}k_i^2$ with solutions $k_i=|k_i|e^{i\chi_i}$, where
\begin{align}\label{eq:k_eps}
|k_i|=\sqrt{\frac{|\sigma_i|}{D_{22}}},\qquad \chi_i=\frac{\mathrm{arg}(\sigma_i)-\pi}{2}+\pi n_i,
\end{align}
and $n_i$ is an integer. Therefore solutions ${w}^i_2$ can both decay and grow exponentially with $r$ on finite domains. On an infinite domain, however, only solutions that decay exponentially with $r$ are allowed ($\mathrm{Re}(k_i) < 0$), which explains exponential localization of the adjoints with a length scale $\ell_-=\mathrm{Re}(k_i)^{-1}\sim\sqrt{D_{22}/|\sigma_i|}$. 
 
\begin{figure}
\includegraphics[width=\columnwidth]{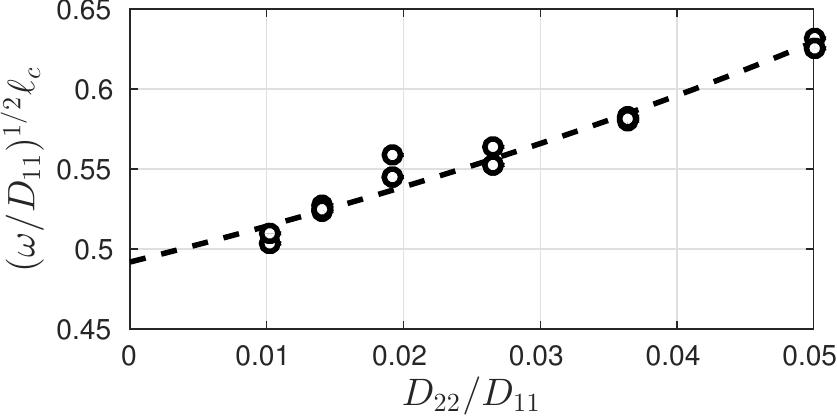} 
\caption{Dependence of the localization length scales $\ell_{c}$ for each response function on $D_{22}$. The dashed line corresponds to a linear fit of $\ell_{c}$ and a quadratic fit of $\omega$ with $D_{22}$.
\label{fig:ell_vs_D22}}
\end{figure}

In the limit $D_{22}\to 0$ (which is the typical case considered in models of cardiac tissue), $D_{11}$ becomes the only parameter in equation \eqref{eq:left} with the dimension of length.
Hence the localization length scale for all slow (unstable, marginal, and weakly stable) modes characterized by the time scale $\omega^{-1}$ can be found using dimensional analysis, which yields $\ell_- \sim\sqrt{D_{11}/\omega} = 5.90$. 
This is fairly close to the numerical value found for the response functions, $\ell_{c}\approx 3.72$, for $D_{22}/D_{11}=0.05$.  Indeed, this is not entirely unexpected: as Fig. \ref{fig:ell_vs_D22} illustrates, the localization length scales for the amplitude of all three marginal adjoint eigenfunctions ($\vec{w}^{x}$, $\vec{w}^{y}$, $\vec{w}^{t}$) depends rather weakly on $D_{22}$.  

For strongly stable modes the relevant time scale can be quite different, although this difference may only become apparent for very quickly decaying modes that are not resolved in the numerics. 
For instance, for $|\Lambda|=0.01$ we have $|\sigma_i|\sim-\ln|\Lambda_i|/T=0.08$ and therefore $\ell_-\sim\sqrt{D_{22}/|\sigma_i|}=1.6$, which is of the same order of magnitude as the value we found for the marginal modes. 
The similarity of the length scales predicted for slow modes and the strongly contracting modes explains the relatively small variation in the localization between adjoint eigenfunctions throughout most of the spectrum.

It is also instructive to compare the structure of the response functions adjoint to the Goldstone modes associated with spatial translations with the structure of the shift map that defines how interaction with a no-flux boundary affects the drift of relative periodic solutions. We have shown previously \cite{Marcotte2015} that the distance $\zeta_n$ between the spiral core and the (planar) boundary after $n$ periods of the revolution can be described by a map
\begin{align}
\zeta^{n+1}=\zeta^n+h_n(\zeta^n).
\end{align}
The roots \rgedit{$0<\zeta_0<\zeta_1<\cdots$ of the shift function $h_n(\zeta)$ define the equilibrium separation values.} When the distance between the origin $\vec{x}_{o}$ of the spiral and the closest boundary is equal to one of these equilibrium values, the spiral wave will drift tangentially to the boundary. An equilibrium $\zeta_k$ is stable provided $|1+h'_n(\zeta_k)|<1$ and unstable otherwise. Since $|h'_n(\zeta)|\ll 1$, this inequality is equivalent to $h'_n(\zeta_k)<0$. In particular, $\zeta_0=5.36\ell_{c}$ is a stable fixed point, while $\zeta_1=7.85\ell_{c}$ is unstable. The existence of stable equilibria suggests the presence of bound states, where a spiral would drift along a planar no-flux boundary forever. Similar bound states were found for resonantly driven spirals next to effective boundaries \cite{LanBar13,LanBar14}.

The drift of spiral waves caused by interaction with the boundaries can be understood by considering the relation between solutions of \eqref{eq:RDE} on bounded domain and on \rgedit{$\mathbb{R}^2$}. 
\rgedit{Consider the flux $\vec{j}=\sigma\nabla\vec{u}$, where the diffusion tensor $\sigma=D$ inside a bounded domain $\Omega$ and $\sigma=0$ outside. In this case, the no-flux boundary condition \eqref{eq:BC} is equivalent to the inclusion of an additional term,
\begin{equation}\label{eq:rde_bound}
	\delta \vec{F}(\vec{u})=\nabla \sigma(\vec{x})\cdot\nabla\vec{u}=-D\delta_{\partial\Omega}(\vec{x})
(\vec{n}\cdot\nabla)\vec{u},
\end{equation}
on the right-hand-side of \eqref{eq:RDE} \rgedit{defined on \rgedit{$\mathbb{R}^2$}}. Here $\delta_{\partial\Omega}$ denotes a one-dimensional delta function localized at $\partial\Omega$, such that in a small neighborhood of every point $\vec{x}_b\in\partial\Omega$}
\begin{equation}\label{eq:delta}
\delta_{\partial\Omega}(\vec{x})=\delta(\vec{n}\cdot(\vec{x}-\vec{x}_b)).
\end{equation}

In the absence of this additional term, \eqref{eq:RDE} possesses a spiral wave solution rigidly rotating around the tip $\vec{x}_o$, which corresponds to the relative equilibrium on \rgedit{$\mathbb{R}^2$}. The introduction of this term generates a perturbation to the dynamics of {\it all} the modes of this solution.
In particular, the perturbation along the Goldstone modes $\vec{v}^x$, $\vec{v}^y$, and $\vec{v}^t$ will generate, respectively, the drift of the spiral core in the $\hat{x}$ and $\hat{y}$ directions and a phase shift (rotation). 
It will be convenient for us to define $\vec{v}^q(t,\vec{x})$ satisfying \eqref{eq:right} such that
\begin{equation}
\vec{v}^q(0,\vec{x})=\vec{v}^q(T,\vec{x})=\partial_q\vec{u}(0,\vec{x}),
\end{equation}
where $q=\{x,y,t\}$ and
\begin{equation}
\vec{w}^q(t,\vec{x})=\sum_{\Lambda_j=1}\alpha_{qj}\vec{w}^j(t,\vec{x}),
\end{equation}
where the coefficients $\alpha_{qj}$ are chosen such that the orthogonality condition $\langle\vec{w}^p|\vec{v}^q\rangle=\delta_{pq}$ is satisfied at $t=T$. 
Thus defined, $\vec{w}^q(t,\vec{x})$ will satisfy \eqref{eq:left} and be orthogonal with respect to $\vec{v}^q(t,\vec{x})$ at all $t$. The spatial drift is then given by~\cite{Biktasheva:2010co},
\begin{equation}\label{eq:h1}
	\dot{\vec{x}}_o(t) = \sum_{q=x,y} \hat{q} \, \langle \vec{w}^q(t) | \delta \vec{F}(\vec{u}(t)) \rangle.
\end{equation}
It should be stressed that this relation is only exact for infinitesimal perturbations, while \eqref{eq:rde_bound} is not infinitesimal. Nonetheless, \eqref{eq:h1} provides a fairly accurate description of the drift, as we will see below.

We can make further progress assuming the boundary $\partial\Omega$ is a smooth curve. With the help of \eqref{eq:rde_bound} the components of \eqref{eq:h1} can be rewritten as
\begin{align}\label{eq:qdot}
	\dot{q}_o(t) = -\int_{\partial\Omega} \dd l[\vec{w}^q(t,\vec{x})]^\dagger 
	D(\vec{n}\cdot\nabla)\vec{u}(t,\vec{x}),
\end{align}
where $\dd l$ is the arclength element along the boundary. Placing the origin of the coordinate system at the tip of the spiral wave, we can write $\vec{w}^q(t,\vec{x})= \hat{\vec{w}}^q(t,\vec{x})e^{-r/\ell_{c}}$, where $r=\sqrt{x^2+y^2}$ and $\hat{\vec{w}}^q(t,\vec{x})=O(1)$.  Since $(\vec{n}\cdot\nabla)\vec{u}(t,\vec{x})=O(1)$ as well, the integral is dominated by the region of the contour around the point $\vec{x}_b$ closest point to origin. Let us orient the coordinate axes such that $\vec{x}_b=(\zeta,0)$ and \cmedit{$\vec{n}=\hat{x}$.} Since on the contour of integration
\begin{align}
	e^{-r/\ell_{c}}\approx e^{-\zeta/\ell_{c}}e^{-y^2/(2\zeta\ell_{c})},
\end{align}
the integral can be evaluated using the saddle point method yielding
\begin{align}\label{eq:h2}
\dot{q}_o(t) \approx -\sqrt{2\pi\zeta\ell_{c}}e^{-\zeta/\ell_{c}}
	[\hat{\vec{w}}^q(t,\vec{x}_b)]^\dagger D\partial_x\vec{u}(t,\vec{x}_b).
\end{align}
Higher order corrections can be easily generated and scale as $(\zeta\ell_c/\lambda^2)^{m}$ relative to \eqref{eq:h2} with integer $m \geq 1$, where $\lambda^2 \gg \zeta\ell_c$, for this spiral wave solution.
Outside of the core, the Archimedian approximation \eqref{eq:Arc} can be used to evaluate the spatial derivative in \eqref{eq:h2}:
\begin{align}
\partial_x\vec{u}(t,\vec{x}_b) = \vec{u}_{0}'(\xi_{b}),
\end{align}
where $\xi_b=\zeta-ct$, if we choose $\theta=0$ on the $x$ axis.

The displacement $\vec{h}=(h_x,h_y)$ can be found by integrating \eqref{eq:h2}, where $\vec{x}_b$ will be a function of $\vec{x}_o$. For a boundary with low curvature ($\kappa\ll\zeta^{-1}$), we only need to keep track of the change in the distance $\zeta$ which satisfies
\begin{equation}\label{eq:zetadot}
\dot{\zeta}=-\dot{x}_o.
\end{equation}
In particular, when $\zeta\gg\ell_{c}$ we can neglect the change in $\zeta$, so that the normal component of the shift $h_n=\cmedit{\vec{n}}\cdot\vec{h}=h_x$ is given by
\begin{align}\label{eq:h3}
	h_n(\zeta) \approx -\sqrt{2\pi\zeta\ell_{c}} e^{-\zeta/\ell_{c}}  \int_{0}^{T} \dd t \, [\hat{\vec{w}}^x(t,\vec{x}_{b})]^{\dagger}D\vec{u}_{0}'(\xi_b).
\end{align}
The dependence of the drift function on the parameters $\epsilon$, $D_{11}$, and $D_{22}$ can be made more explicit by factoring out the dependence of the components of the solution and the response function on the small parameter $\epsilon$: $\hat{\vec{w}}^x = (\epsilon\tilde{\vec{w}}^x_1,\tilde{\vec{w}}^x_2)$ and  $\vec{u}'_{0} = (\tilde{\vec{u}}'_{0,1},\epsilon\tilde{\vec{u}}'_{0,2})$. Then \eqref{eq:h3} can be rewritten as
\begin{align}
h_n(\zeta) \approx \ell_d\bar{h}_n(\zeta) e^{-\zeta/\ell_{c}},
\end{align}
where 
\begin{align}\label{eq:Iz}
 \bar{h}_n(\zeta)\equiv-\frac{c\ell_r}{\tr D} \sqrt{\frac{\zeta}{\ell_{c}}}\int_0^T \dd t \, [\tilde{\vec{w}}^x(t,\vec{x}_{b})]^\dagger D \tilde{\vec{u}}'_0(\zeta - ct)
\end{align}
is a non-dimensional function,
\begin{align}\label{eq:h0}
\ell_d \equiv \sqrt{2\pi}\frac{\epsilon \ell_{c}}{\ell_r} \frac{\tr D}{c},
\end{align}
and $\ell_r$ are characteristic length scales that represents the dependence of the amplitude of the response function $\vec{w}^x$ on parameters.

Rather predictably, we find that the dominant contribution to the drift is determined by the component of the state which has the largest diffusion constant. In the Karma model considered here $D_{11} \gg D_{22}$, and so it is the first component (which corresponds to the voltage variable) that controls the drift. For the inner product generally, where the contributions from the first and second components are unweighted, it is the contribution of the second component which dominates. Hence the normalization condition for $\vec{w}^x$ dictates that $\tilde{w}^x\propto \ell_f/\ell_c^2$, where $\ell_f$ is the length scale on which the second component of $\vec{u}_0$ varies. Using the Karma model \eqref{eq:RDE} it is straightforward to show that $\ell_f = \lambda/(2\epsilon)$ (cf. the spatial Goldstone modes in Fig.~\ref{fig:GM+RF}(d-e) and the second component of the solution shown in Fig.~\ref{fig:spectrum}(a)). Therefore, we can define $\ell_r=\ell_c^2/\ell_f=2\epsilon\ell_c^2/\lambda$ and
\begin{align}
\ell_d \equiv \sqrt{\frac{\pi}{2}}\frac{\lambda}{\ell_{c}}\frac{\tr D}{c}.
\end{align}
This \rgedit{length scale} is quite different from a na\"ive guess $\ell_d\sim\ell_c$ \rgedit{based solely on the scaling of the response functions.}

\begin{table}[b!]
\begin{tabular}{|l|c|c|c|}
\hline
 & $A$ & \ $\Delta\zeta/\Delta r$ \ & $\ell_c$ \\ 
\hline 
\ Direct numerical simulation \ & \ -0.430 \ & 0.873 & \ 3.72 \ \\ 
\hline 
\ Integrated drift equation \ & -0.580 & 0.905 & 3.80 \\ 
\hline 
\ Saddle point approximation \ & -0.724 & 0.922 & 3.80 \\ 
\hline 
\end{tabular} 
\caption{Fitting parameters for Eq. \eqref{eq:h_n_gen}.\label{tab:h_params}}
\end{table}

\begin{figure}
\includegraphics[width=\columnwidth]{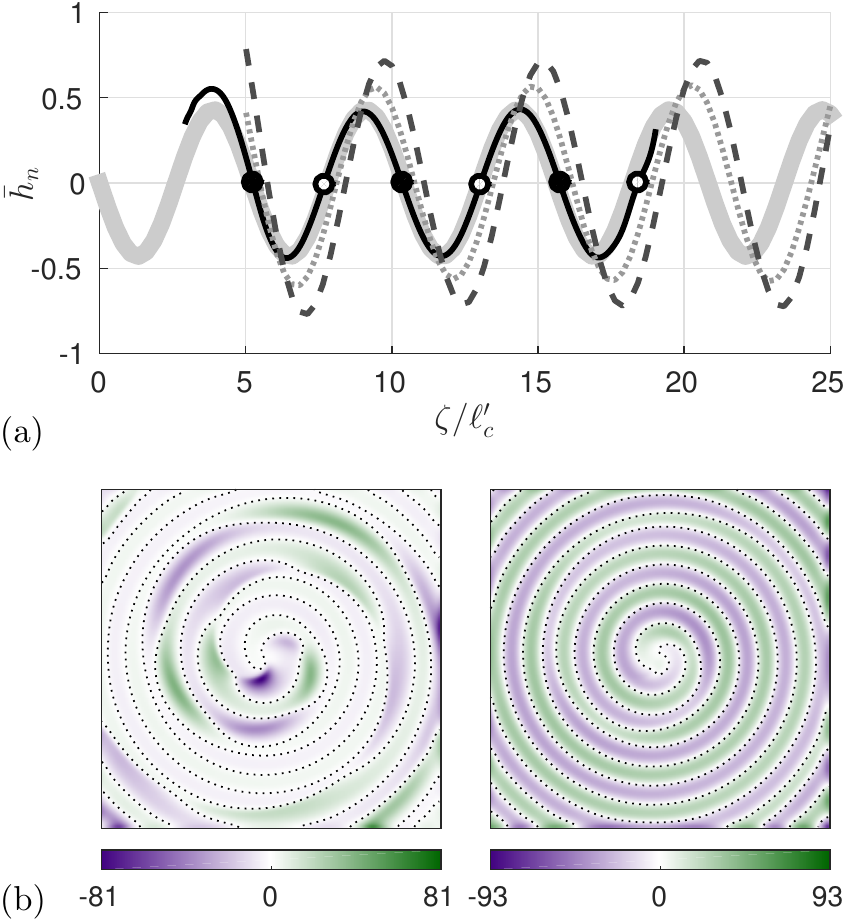} 
\caption{(a) The scaled shift function $\bar{h}_n = (h_n/\ell_d) e^{\zeta/\ell_{c}'}$ obtained by direct numerical simulation (black line) and its fit $A\sin(\pi\zeta/\Delta\zeta)$ (gray). The roots $\zeta_{k}$ are denoted by circles (filled stable, open unstable). The dotted and dashed lines corresponds to the integral of \eqref{eq:h1} over one temporal period and its saddle point approximation \eqref{eq:h3}, respectively. 
(b) Snapshots of the first and second component of the scaled response function $\tilde{\vec{w}}^{x}(t,\vec{x})$ at time $t=T/2$ with nodal lines denoted by dotted curves.
\label{fig:h_x_map}}
\end{figure} 

As Fig. \ref{fig:h_x_map}(b) shows, both components of $\tilde{\vec{w}}^x$ exhibit pronounced oscillatory dependence on the distance $r$ from the origin. The scaled shift function $\bar{h}_n(\zeta)$ inherits this oscillatory dependence: the expression \eqref{eq:Iz} is shown as the dashed line in Fig. \ref{fig:h_x_map}(a).
The amplitude of the oscillation is nearly constant for $\zeta/\ell_c \gg 1$, and to a high accuracy we can fit $\bar{h}_n(\zeta) = A\sin(\pi \zeta/\Delta\zeta)$ with constant $A$ and $\Delta\zeta$.
The corresponding scaling for the shift function
\begin{equation}\label{eq:h_n_gen}
	h_n(\zeta) = \ell_d A \sin(\pi \zeta/\Delta\zeta) e^{-\zeta/\ell_{c}}.
\end{equation}
yields a prediction which is in good agreement with both our previous numerical results \cite{Marcotte2015} and with the drift equation \eqref{eq:qdot} integrated over one temporal period (cf. Fig.~\ref{fig:h_x_map}(a)). The fitting parameters for all three cases are given in Table \ref{tab:h_params}. In particular, we find that the values of $A$ are $O(1)$, which supports our choice of the length scale $\ell_r$. 
The values of $\ell_c$ found by fitting the shifts correspond reasonably closely to the localization length scale of the response function. Moreover, the spacing $\Delta\zeta$ between the roots of the shift function corresponds well to the distance $\Delta r$ between the nodal lines of the response function.
This confirms our conjecture \cite{Marcotte2015} that the interaction of spiral waves with a physical no-flux boundary is controlled by the response functions, just as in the case of resonantly driven spirals interacting with effective boundaries formed by a step-wise change in the excitability of the medium \cite{LanBar13,LanBar14}.

The saddle point approximation noticeably overestimates the magnitude of the shift due to interaction of the spiral wave with the boundary, while integrating \eqref{eq:qdot} directly produces an estimate that is in reasonable quantitative agreement with the result of direct numerical simulations. 
While the saddle point approximation can be easily improved at larger $\zeta/\ell_c$ by retaining higher order terms in $\zeta$, its main value is in uncovering the explicit dependence of the drift on various parameters of the problem.
In practice, it is only the first root of the drift map, $\zeta_{0}$, which is likely to play any role in the dynamics of spiral waves. The rest of the equilibria are essentially marginally stable, and the shift becomes exponentially small. 
The saddle point approximation produces a reasonably good prediction for the value of $\zeta_{0}$. Even the prediction of the spacing $\Delta\zeta$ between the equilibria is in fairly good agreement: the error is only about $0.0485\zeta_{0}$.

\rgedit{Qualitatively similar results were obtained for spiral interaction in CGLE using the amplitude equation formalism \cite{aranson1993interaction} and for spiral interaction with domain boundaries and defects in excitable systems using the kinematic theory \cite{aranson1995drift}. In the latter case, however, the interaction strength was predicted to decay super-exponentially fast (i.e., as $\exp(-\zeta^{3})$), a scaling our results do not support.}

\section{Conclusions \label{sec:c}}

On domains of arbitrary shape, pinned and drifting spiral wave solutions of excitable systems are described, respectively, by temporally periodic and generalized relative periodic solutions. 
We have developed a general numerical procedure that allows computation of the leading adjoint eigenfunctions for unstable spiral wave solutions of such types. In particular, we computed hundreds of the dominant adjoint eigenfunctions for (slowly) drifting single-spiral wave solutions of the Karma model.

Just like for spiral wave solutions described by relative equilibria on circular domains, we found that the response functions, or adjoint eigenfunctions that correspond to marginal degrees of freedom, are exponentially localized in the vicinity of the spiral tip. The localization length scale of the response functions found numerically is in good agreement with the order-of-magnitude estimate $\ell_{c}\sim\sqrt{D_{11}/\omega}$ based on dimensional analysis, where $D_{11}$ is the diffusion constant associated with the {\it fast} variable and $\omega$ is the angular frequency of the underlying spiral wave solution. 

Adjoint eigenfunctions associated with other leading modes, both stable and unstable ones, were also found to be exponentially localized in the vicinity of the spiral tip, with the corresponding localization length scale $\ell_{-}$ larger than $\ell_{c}$. For strongly stable modes it can be shown more rigorously that $\ell_{-}\sim\sqrt{D_{22}/|\sigma|}$, where $D_{22}$ is the diffusion constant associated with the {\it slow} variable and $\sigma$ is the corresponding Floquet exponent.

The significance of response functions for the dynamics of \rgedit{isolated spiral waves on $\mathbb{R}^2$} is well understood\cmedit{\cite{BiHoBi06,BBBBF09,biktasheva2001}}. The spatial and temporal response functions determine the effect of small perturbations in the initial conditions or the evolution equations on, respectively, the drift of the spiral and its rotation speed\cmedit{\cite{Biktasheva:2010co,biktasheva2015}}. 
\rgedit{In particular, the response functions have been used to describe the interaction of spiral waves with tissue heterogeneties\cite{Biktashev2010,Biktashev2011,biktasheva2000drift}.}
Our results further show that the spatial response functions also determine the interaction of spiral waves with physical no-flux boundaries and, by extension, the interaction with neighboring spirals through tile boundaries with effective no-flux boundary conditions \cite{ByMaGr14,Marcotte2015}. Specifically, the spatial response functions $\vec{w}^x$ and $\vec{w}^y$ define the shift function $\vec{h}(\zeta)$ which describes the displacement of the spiral wave origin due to the interaction with the boundary over one temporal period.

The rest of the adjoint eigenfunctions have received very little attention in the literature dealing with excitable systems in general and the dynamics of cardiac tissue in particular. The spatiotemporal structure of unstable and weakly stable adjoints, however, is critically important for understanding spiral wave breakup and chaotic dynamics featuring multiple interacting unstable spiral waves~\cite{Marcotte2016b}. 
For instance, it is well known that stable spirals whose cores are sufficiently well separated can be considered effectively independent. The same is also true of unstable spirals \cite{Marcotte2015}. The spirals begin to interact at smaller separations, with interaction that can be conveniently described with the help of the adjoint eigenfunctions. 
Our results show that not only the position of the core and the phase of a spiral wave, but also its stability should be affected by neighboring spirals.

Furthermore, adjoints associated with slow modes play a crucial role in the design of feedback control methods aimed at suppressing spiral wave instabilities (such as alternans) \cite{Otani2004}. The spatial localization of the adjoints associated with unstable modes indicates that feedback is most effective when it is applied close to the core of a spiral wave. Furthermore, the spatial alternation of the phase of the adjoints associated with all slow modes (unstable, marginal, and weakly stable) in the core region  significantly attenuates the effectiveness of spatially uniform perturbations \cite{agladze1987, biktashev1993resonant, LanBar13} on the dynamics of spiral waves. This suggests that spatially localized perturbation, e.g., those due to virtual electrodes \cite{efimov2000,Fenton2009} should be much more effective for control of spatiotemporally chaotic regimes, such as fibrillation.

\section{Supplementary Material \label{sec:sm}}
	See supplementary material for the temporal evolution of the leading eigenfunction pairs, as well as the rescaled response function used in the computation of the drift.

\begin{acknowledgments}
This material is based upon work supported by the National Science Foundation under Grant No. CMMI-1028133.  
The Tesla K20 GPUs used for this research were donated by the ``NVIDIA Corporation'' through the academic hardware donation program.
\end{acknowledgments}

\section*{Appendix}

\rgedit{The right eigenfunctions of the time-evolution operator $\mathcal{V}_T$ have been computed via Arnoldi iteration \cite{Arnoldi1951} which involves time-integration of the linearized equations \eqref{eq:right}.}
It should be pointed out that, since the instantaneous Jacobian $L$ is a function of the reference state $\vec{u}(t)$, both \eqref{eq:RDE} and \eqref{eq:right} are time-integrated simultaneously to avoid storing and retrieving the reference solution. 
The same spatial (2nd order finite difference on a square mesh) and temporal (4th order fully explicit Runge-Kutta) discretization scheme is used for both equations \cite{Marcotte2015}.

However, the same approach cannot be used to compute the left eigenfunctions, since \eqref{eq:left} should be integrated backwards in time and \eqref{eq:RDE} cannot be time-integrated in the reverse direction. 
Furthermore, as the evolution equation is quite stiff,  fairly small time steps have to be used ($O(10^4)$ time steps per period). One period of a fully resolved solution corresponds to about 8 GB of data, which may not fit in RAM.
Hence, the entire reference solution $\vec{u}(t)$ must be pre-computed, stored, and then retrieved during the time-integration of \eqref{eq:left}. 

\cmedit{The spatial discretization and the time-stepping of the adjoint tangent evolution \eqref{eq:left} are the same as those for \eqref{eq:RDE} and \eqref{eq:right}.}
This choice was made out of necessity: the discrete adjoint of an explicit Runge-Kutta method is at least semi-implicit \cite{hager2000runge,sandu2006properties}.
For a partial-differential equation, this requires solution of an infeasibly large linear system at every time step.
Furthermore, the discretization of \eqref{eq:left} is sufficiently precise that the solution of the large linear system is unnecessary.

\begin{figure}[htpb]
	\includegraphics[width=\columnwidth]{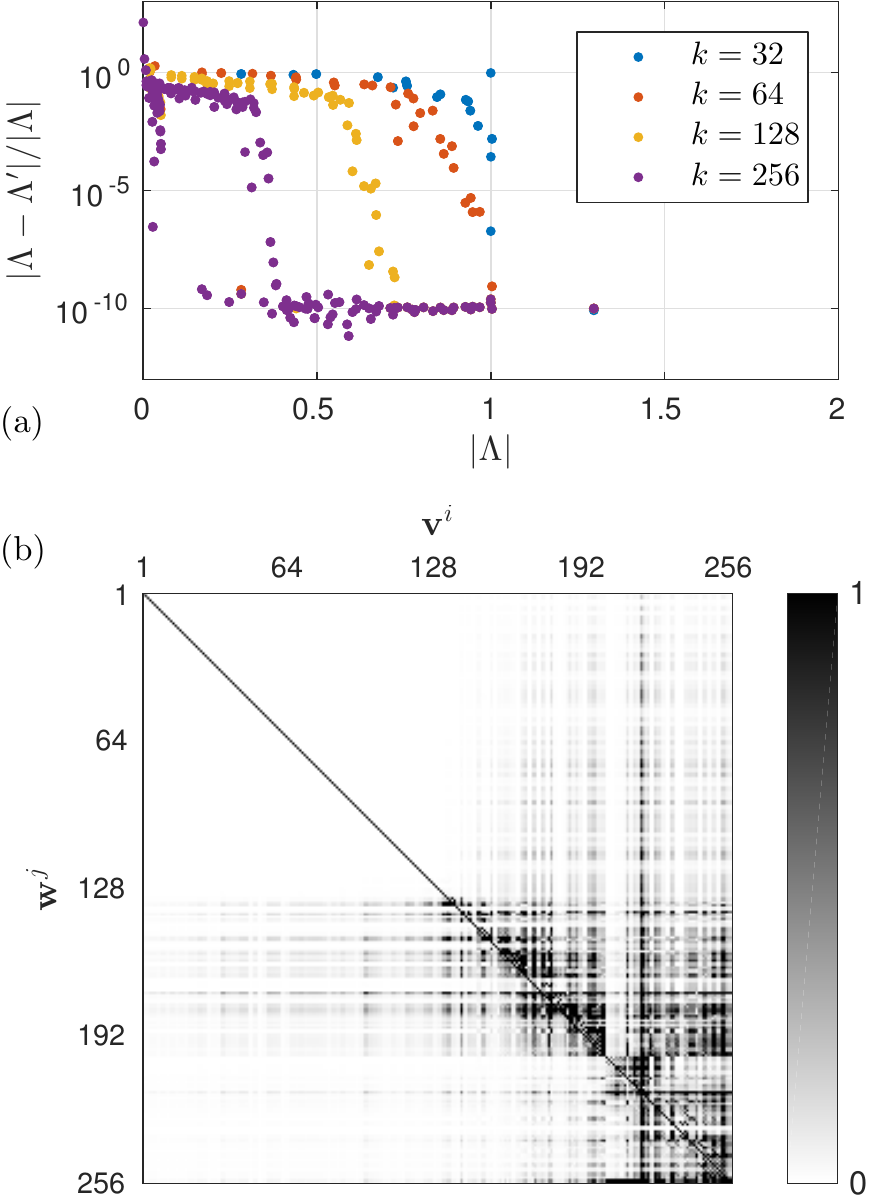}
\caption{ (a) Relative eigenvalue deviations $|\Lambda - \Lambda^{\prime}|/|\Lambda|$ from the leading $k$-dimensional Krylov subspace.
	(b) The inner product of the leading set of left ($\vec{w}^{j}$) and right ($\vec{v}^{i}$)  eigenfunctions.
\label{fig:eigerr}}
\end{figure}

\rgedit{Runge-Kutta integrators for \eqref{eq:left} require evaluation of $\vec{u}(t)$ at intermediate points between the time steps, while the solution $\vec{u}^n=\vec{u}(t_n)$ is only known at discrete times $t_n=n\Delta t$. 
To preserve the accuracy of time-integration of \eqref{eq:left}, we use a high-order interpolation of $\vec{u}(t)$. 
That is, for an integrator of order $O(\Delta t^p)$, we use an interpolant uniformly accurate on the interval $t\in [t_n,t_{n+1}]$ to order $O(\Delta t^q)$, with $q \geq p$.}
Following the methodology of Enright {\em et. al.}~\cite{enright1986interpolants}, the $O(\Delta t^4)$ interpolant for the classical Runge-Kutta method used in this work is 
\begin{align}
	\vec{u}_4(t + \tau \Delta t) 
		&= d_{4,0}(\tau) \vec{u}^{n} 
			+ d_{4,1}(\tau) \vec{u}^{n+1} \\
		&+ d_{4,2}(\tau) \partial_t\vec{u}^{n}\Delta t
			+ d_{4,3}(\tau) \partial_t\vec{u}^{n+1}\Delta t\ \nonumber \\
		&+ d_{4,4}(\tau) \partial_t\vec{u}_{3}(t + \eta \Delta t)\Delta t\ 
			+ O(\Delta t^{5}),\nonumber 
\end{align}
where $\vec{u}_{3}(t + \eta \Delta t)$ is the value of the $O(\Delta t^3)$ interpolated state at time $t_{n} + \eta \Delta t$, which for this method has $\eta=1/3$.
The coefficients for the fourth order interpolant  are
\begin{align}\label{eq:d4}
	d_{4,0}(\tau) &= 1 + 6 \tau^2 - 16 \tau^3 + 9\tau^4, \nonumber\\
	d_{4,2}(\tau) &= \tau - 2 \tau^2 + \tau^3, 			\nonumber\\
	d_{4,3}(\tau) &= \left(5\tau^2 - 14\tau^3 + 9\tau^4\right)/4,	\nonumber\\
	d_{4,4}(\tau) &= 27\left(\tau^2 - 2\tau^3 +  \tau^4 \right)/4,	
\end{align}
where $d_{4,1}(\tau) = 1-d_{4,0}(\tau)$.
\rgedit{Due to interpolation, it takes roughly \cmedit{twice as long} to integrate the adjoint evolution equation \eqref{eq:left} compared with the tangent evolution equation \eqref{eq:right} (\cmedit{e.g.,} 31 hours vs. 70 hours to compute the spectra shown in Fig. \ref{fig:eigerr} on a single NVIDIA Tesla K20 GPU). }

\rgedit{The Arnoldi iteration procedure we used to compute the spectrum (eigenfunctions and eigenvalues) of the finite-time tangent evolution map $\mathcal{V}_T$ can be applied to any time-periodic solution $\vec{u}(t)$. It relies on time-integration of \eqref{eq:right} forward in time (with the final state $\vec{v}_i(T)$ for one iteration used to define the initial condition $\vec{v}_{i+1}(0)$ for the next iteration), to} generate a sufficiently large orthonormal basis $\{\hat{\vec{v}}_1,\cdots,\hat{\vec{v}}_k\}$ which defines the Krylov subspace that contains the dominant modes and an approximate Jacobian $H_k$ in that basis 
\begin{equation}
	\mathcal{V}_T{\bf\hat{v}}_{k} \approx H_{k}{\bf\hat{v}}_{k}.
\end{equation}

The adjoint spectrum  (eigenfunctions and eigenvalues of $\mathcal{V}_T^\dagger$) is computed using \rgedit{a similar approach (time-integrating \eqref{eq:left} backwards in time with the final state $\vec{w}_i(0)$ for one iteration used to define the initial condition $\vec{w}_{i+1}(T)$ for the next iteration)}, yielding an orthonormal basis $\{\hat{\vec{w}}_1,\cdots,\hat{\vec{w}}_k\}$ and an approximate Jacobian $H'_k$
\begin{equation}
	\mathcal{V}_T^{\dagger}{\bf\hat{w}}_{k} \approx H_{k}' {\bf\hat{w}}_{k}.
\end{equation}
The eigenvalues and eigenvectors of the matrices $H_k$, $H'_k$ approximate the eigenvalues and (projected) eigenfunctions of the time-evolution operator and its adjoint, respectively.

The \cmedit{generation of eigenfunctions via the} Arnoldi iteration relies on the \cmedit{closure} of the reference solution $\vec{u}(t)$, \cmedit{as iteration forward in time by $\mathcal{V}_T$ requires both the beginning and endpoints of the trajectory to coincide in state-space}.
\cmedit{If the beginning and end states are distinct, then so are the associated tangent spaces, and successive applications of the map $\mathcal{V}_T$} \rgedit{become ill-defined.}
The set of basis vectors that defines the Krylov subspace is constructed by performing Gram-Schmidt orthogonalization of the sequence of \cmedit{Krylov vectors $\vec{a}_k$ -- collocated in state space about the initial condition $\vec{u}(0)$ -- } constructed using the iterative relation
\begin{equation}\label{eq:arn_po}
\vec{a}_k=\mathcal{V}_T\vec{a}_{k-1},
\end{equation}
which span the dominant subspace of the operator. For relative periodic orbits $\vec{u}(T)=\mathcal{U}_T\vec{u}(0)=g^{-1}\vec{u}(0)$, where $g\ne 1$ is a symmetry transformation. This allows generalization of the iterative relation
\begin{equation}\label{eq:arn_rpo}
\vec{a}_k=g\mathcal{V}_T\vec{a}_{k-1},
\end{equation}
again yielding a collocated set of Krylov vectors that can be used to construct an orthonormal basis.

For open trajectories\rgedit{, where the beginning and end points (and the respective tangent spaces) do not coincide even after a symmetry transformation,} this is no longer an option, since there is no group action that can be used to \rgedit{reinitialize the iteration}. 
Instead a collocated basis set must be formed at each end of the trajectory: $\{\vec{a}_1,\cdots,\vec{a}_k\}$ at $\vec{u}(0)$ and $\{\vec{b}_1,\cdots,\vec{b}_k\}$ at $\vec{u}(T)$, where 
\begin{align}
\vec{b}_k&=\mathcal{V}_T\vec{a}_k,\nonumber\\
\vec{a}_{k+1}&=\mathcal{V}_T^\dagger\vec{b}_k.
\end{align}
The basis sets $\{\hat{\vec{v}}_1,\cdots,\hat{\vec{v}}_k\}$ and $\{\hat{\vec{w}}_1,\cdots,\hat{\vec{w}}_k\}$ can then be found by orthogonalizing the sets $\{\vec{a}_1,\cdots,\vec{a}_k\}$ and $\{\vec{b}_1,\cdots,\vec{b}_k\}$.
\rgedit{The difference in the two approaches reflect the fact that while for (relative) periodic solutions the tangent evolution operator is most conveniently represented using its spectral decomposition, for open trajectories the relevant representation is based on the singular value decomposition.}

\cmedit{While the accuracy of the nonlinear map $\mathcal{U}_T$ and forward-tangent evolution map $\mathcal{V}_T$ can be assessed using the typical methods of convergence analysis, the tools available to verify the accuracy of the adjoint tangent evolution $\mathcal{V}_T^{\dagger}$ are limited.}
The accuracy of the backwards time integration can not be assessed in the typical way by increasing the temporal and spatial resolution of the discretization of \eqref{eq:left} since these are set by the discretization of the nonlinear equation \eqref{eq:RDE}, \cmedit{and thus can not be modified independently of the forward-time solution}.
However, there are two independent parameters which we can vary: the order $p$ of the interpolation method and the order $q$ of the integration scheme used for the adjoint time-stepping.
Our results suggest that the error in computing the adjoints is dominated by the interpolation order when $q\leq 3$, with significant increases in accuracy when both $q=4$ and $p=4$.
As discussed below, the reliability of the adjoint tangent evolution may be indirectly measured either by the magnitude of the inner product \eqref{eq:biorth} for $i\ne j$ or by the difference between the eigenvalues $\Lambda_i$ of $H_k$ and the eigenvalues $\Lambda'_i$ of $H'_k$.
As Fig. \ref{fig:eigerr} illustrates, a 256-dimensional Krylov subspace allows computing $\sim$130 leading modes with \cmedit{high} accuracy. 
More generally, a $k$-dimensional Krylov subspace allows accurate determination of up to $k/2$ modes with the relative error $|\Lambda' - \Lambda|/|\Lambda|=O(10^{-10})$.

As the computation of the left and right eigenfunctions involves two distinct matrices representing compact truncations of the formally infinite-dimensional evolution operator and its adjoint, there is some ambiguity in matching these two sets of eigenfunctions. 
The eigenfunctions can be matched based on the closeness of the associated eigenvalues, which is the choice made in the present paper. 
The right eigenvalues are ordered by their absolute value, from the largest to the smallest $|\Lambda_1|\geq|\Lambda_2|\geq\cdots\geq|\Lambda_k|$.
Each right eigenvalue $\Lambda_i$ and the corresponding eigenfunction $\vec{v}^{i}$ is then paired with the left eigenvalue $\Lambda'_j$ and eigenfunction $\vec{w}^{j}$, such that $j\geq i$ corresponds to the smallest value of $|\Lambda'_j - \Lambda_i|$. 
This matching procedure makes no assumption regarding the orthogonality between the two eigenfunction sets, so the condition \eqref{eq:biorth} can be used to check the accuracy with which the eigenfunctions have been computed. 

Alternatively, the eigenfunctions can be matched based on the orthogonality relation \eqref{eq:biorth}.
In this case each left eigenfunction $\vec{w}^{i}$ and the corresponding left eigenvalue $\Lambda_i'$ is matched with the right eigenfunction $\vec{v}^{j}$ and eigenvalue $\Lambda_j$, such that $j\geq i$ corresponds to the largest value of the inner product $\langle\vec{w}^{i}|\vec{v}^{j}\rangle$, where both sets have been independently normalized to unity, beforehand.
This procedure has the benefit of most closely reproducing the orthogonality condition \eqref{eq:biorth}. 
The differences $|\Lambda_i-\Lambda'_i|$ can be used to assess the accuracy with which the eigenvalues have been computed. 
\cmedit{Both methods of matching the right and left sets yield the same results for the resolved eigenmodes (that is, those which are effectively captured by a sufficiently large Krylov space, or the leading $k/2$ modes in our $k=256$-dimensional space).}

\section*{References}
\bibliography{../bibtex/cardiac}

\begin{thebibliography}{54}%
\makeatletter
\providecommand \@ifxundefined [1]{%
 \@ifx{#1\undefined}
}%
\providecommand \@ifnum [1]{%
 \ifnum #1\expandafter \@firstoftwo
 \else \expandafter \@secondoftwo
 \fi
}%
\providecommand \@ifx [1]{%
 \ifx #1\expandafter \@firstoftwo
 \else \expandafter \@secondoftwo
 \fi
}%
\providecommand \natexlab [1]{#1}%
\providecommand \enquote  [1]{``#1''}%
\providecommand \bibnamefont  [1]{#1}%
\providecommand \bibfnamefont [1]{#1}%
\providecommand \citenamefont [1]{#1}%
\providecommand \href@noop [0]{\@secondoftwo}%
\providecommand \href [0]{\begingroup \@sanitize@url \@href}%
\providecommand \@href[1]{\@@startlink{#1}\@@href}%
\providecommand \@@href[1]{\endgroup#1\@@endlink}%
\providecommand \@sanitize@url [0]{\catcode `\\12\catcode `\$12\catcode
  `\&12\catcode `\#12\catcode `\^12\catcode `\_12\catcode `\%12\relax}%
\providecommand \@@startlink[1]{}%
\providecommand \@@endlink[0]{}%
\providecommand \url  [0]{\begingroup\@sanitize@url \@url }%
\providecommand \@url [1]{\endgroup\@href {#1}{\urlprefix }}%
\providecommand \urlprefix  [0]{URL }%
\providecommand \Eprint [0]{\href }%
\providecommand \doibase [0]{http://dx.doi.org/}%
\providecommand \selectlanguage [0]{\@gobble}%
\providecommand \bibinfo  [0]{\@secondoftwo}%
\providecommand \bibfield  [0]{\@secondoftwo}%
\providecommand \translation [1]{[#1]}%
\providecommand \BibitemOpen [0]{}%
\providecommand \bibitemStop [0]{}%
\providecommand \bibitemNoStop [0]{.\EOS\space}%
\providecommand \EOS [0]{\spacefactor3000\relax}%
\providecommand \BibitemShut  [1]{\csname bibitem#1\endcsname}%
\let\auto@bib@innerbib\@empty
\bibitem [{\citenamefont {Barkley}(1994)}]{Barkley94}%
  \BibitemOpen
  \bibfield  {author} {\bibinfo {author} {\bibfnamefont {D.}~\bibnamefont
  {Barkley}},\ }\bibfield  {title} {\enquote {\bibinfo {title} {{Euclid}ean
  symmetry and the dynamics of rotating spiral waves},}\ }\href@noop {}
  {\bibfield  {journal} {\bibinfo  {journal} {Phys. Rev. Lett.}\ }\textbf
  {\bibinfo {volume} {72}},\ \bibinfo {pages} {164--167} (\bibinfo {year}
  {1994})}\BibitemShut {NoStop}%
\bibitem [{\citenamefont {Biktasheva}, \citenamefont {Elkin},\ and\
  \citenamefont {Biktashev}(1998)}]{biktasheva1998}%
  \BibitemOpen
  \bibfield  {author} {\bibinfo {author} {\bibfnamefont {I.~V.}\ \bibnamefont
  {Biktasheva}}, \bibinfo {author} {\bibfnamefont {Y.~E.}\ \bibnamefont
  {Elkin}}, \ and\ \bibinfo {author} {\bibfnamefont {V.~N.}\ \bibnamefont
  {Biktashev}},\ }\bibfield  {title} {\enquote {\bibinfo {title} {Localized
  sensitivity of spiral waves in the complex {Ginzburg-Landau} equation},}\
  }\href@noop {} {\bibfield  {journal} {\bibinfo  {journal} {Phys. Rev. E}\
  }\textbf {\bibinfo {volume} {57}},\ \bibinfo {pages} {2656--2659} (\bibinfo
  {year} {1998})}\BibitemShut {NoStop}%
\bibitem [{\citenamefont {Keener}(1988)}]{keener1988dynamics}%
  \BibitemOpen
  \bibfield  {author} {\bibinfo {author} {\bibfnamefont {J.~P.}\ \bibnamefont
  {Keener}},\ }\bibfield  {title} {\enquote {\bibinfo {title} {The dynamics of
  three-dimensional scroll waves in excitable media},}\ }\href@noop {}
  {\bibfield  {journal} {\bibinfo  {journal} {Physica D}\ }\textbf {\bibinfo
  {volume} {31}},\ \bibinfo {pages} {269--276} (\bibinfo {year}
  {1988})}\BibitemShut {NoStop}%
\bibitem [{\citenamefont {Biktashev}\ and\ \citenamefont
  {Holden}(1993)}]{biktashev1993resonant}%
  \BibitemOpen
  \bibfield  {author} {\bibinfo {author} {\bibfnamefont {V.}~\bibnamefont
  {Biktashev}}\ and\ \bibinfo {author} {\bibfnamefont {A.}~\bibnamefont
  {Holden}},\ }\bibfield  {title} {\enquote {\bibinfo {title} {Resonant drift
  of an autowave vortex in a bounded medium},}\ }\href@noop {} {\bibfield
  {journal} {\bibinfo  {journal} {Physics Letters A}\ }\textbf {\bibinfo
  {volume} {181}},\ \bibinfo {pages} {216--224} (\bibinfo {year}
  {1993})}\BibitemShut {NoStop}%
\bibitem [{\citenamefont {Biktashev}\ and\ \citenamefont
  {Holden}(1995)}]{Biktashev:1995re}%
  \BibitemOpen
  \bibfield  {author} {\bibinfo {author} {\bibfnamefont {V.~N.}\ \bibnamefont
  {Biktashev}}\ and\ \bibinfo {author} {\bibfnamefont {A.~V.}\ \bibnamefont
  {Holden}},\ }\bibfield  {title} {\enquote {\bibinfo {title} {Resonant drift
  of autowave vortices in two dimensions and the effects of boundaries and
  inhomogeneities},}\ }\href@noop {} {\bibfield  {journal} {\bibinfo  {journal}
  {Chaos Soliton Fract.}\ }\textbf {\bibinfo {volume} {5}},\ \bibinfo {pages}
  {575--622} (\bibinfo {year} {1995})}\BibitemShut {NoStop}%
\bibitem [{\citenamefont {Henry}\ and\ \citenamefont
  {Hakim}(2000)}]{Henry2000}%
  \BibitemOpen
  \bibfield  {author} {\bibinfo {author} {\bibfnamefont {H.}~\bibnamefont
  {Henry}}\ and\ \bibinfo {author} {\bibfnamefont {V.}~\bibnamefont {Hakim}},\
  }\bibfield  {title} {\enquote {\bibinfo {title} {Linear stability of scroll
  waves},}\ }\href@noop {} {\bibfield  {journal} {\bibinfo  {journal} {Phys.
  Rev. Lett.}\ }\textbf {\bibinfo {volume} {85}},\ \bibinfo {pages}
  {5328--5331} (\bibinfo {year} {2000})}\BibitemShut {NoStop}%
\bibitem [{\citenamefont {Henry}\ and\ \citenamefont
  {Hakim}(2002)}]{Henry2002}%
  \BibitemOpen
  \bibfield  {author} {\bibinfo {author} {\bibfnamefont {H.}~\bibnamefont
  {Henry}}\ and\ \bibinfo {author} {\bibfnamefont {V.}~\bibnamefont {Hakim}},\
  }\bibfield  {title} {\enquote {\bibinfo {title} {Scroll waves in isotropic
  excitable media: Linear instabilities, bifurcations, and restabilized
  states},}\ }\href@noop {} {\bibfield  {journal} {\bibinfo  {journal} {Phys.
  Rev. E}\ }\textbf {\bibinfo {volume} {65}},\ \bibinfo {pages} {046235}
  (\bibinfo {year} {2002})}\BibitemShut {NoStop}%
\bibitem [{\citenamefont {Biktasheva}, \citenamefont {Elkin},\ and\
  \citenamefont {Biktashev}(1999)}]{biktasheva1999resonant}%
  \BibitemOpen
  \bibfield  {author} {\bibinfo {author} {\bibfnamefont {I.~V.}\ \bibnamefont
  {Biktasheva}}, \bibinfo {author} {\bibfnamefont {Y.~E.}\ \bibnamefont
  {Elkin}}, \ and\ \bibinfo {author} {\bibfnamefont {V.~N.}\ \bibnamefont
  {Biktashev}},\ }\bibfield  {title} {\enquote {\bibinfo {title} {Resonant
  drift of spiral waves in the complex {Ginzburg-Landau} equation},}\
  }\href@noop {} {\bibfield  {journal} {\bibinfo  {journal} {Journal of
  biological physics}\ }\textbf {\bibinfo {volume} {25}},\ \bibinfo {pages}
  {115--127} (\bibinfo {year} {1999})}\BibitemShut {NoStop}%
\bibitem [{\citenamefont {Biktasheva}\ and\ \citenamefont
  {Biktashev}(2001)}]{biktasheva2001}%
  \BibitemOpen
  \bibfield  {author} {\bibinfo {author} {\bibfnamefont {I.~V.}\ \bibnamefont
  {Biktasheva}}\ and\ \bibinfo {author} {\bibfnamefont {V.~N.}\ \bibnamefont
  {Biktashev}},\ }\bibfield  {title} {\enquote {\bibinfo {title} {Response
  functions of spiral wave solutions of the complex {G}inzburg-{L}andau
  equation},}\ }\href@noop {} {\bibfield  {journal} {\bibinfo  {journal} {J.
  Nonlin. Math. Phys.}\ }\textbf {\bibinfo {volume} {8}},\ \bibinfo {pages}
  {28--34} (\bibinfo {year} {2001})}\BibitemShut {NoStop}%
\bibitem [{\citenamefont {Aranson}, \citenamefont {Kramer},\ and\ \citenamefont
  {Weber}(1991)}]{aranson1991interaction}%
  \BibitemOpen
  \bibfield  {author} {\bibinfo {author} {\bibfnamefont {I.}~\bibnamefont
  {Aranson}}, \bibinfo {author} {\bibfnamefont {L.}~\bibnamefont {Kramer}}, \
  and\ \bibinfo {author} {\bibfnamefont {A.}~\bibnamefont {Weber}},\ }\bibfield
   {title} {\enquote {\bibinfo {title} {On the interaction of spiral waves in
  non-equilibrium media},}\ }\href@noop {} {\bibfield  {journal} {\bibinfo
  {journal} {Physica D}\ }\textbf {\bibinfo {volume} {53}},\ \bibinfo {pages}
  {376--384} (\bibinfo {year} {1991})}\BibitemShut {NoStop}%
\bibitem [{\citenamefont {Pismen}\ and\ \citenamefont
  {Nepomnyashchy}(1991)}]{pismen1991mobility}%
  \BibitemOpen
  \bibfield  {author} {\bibinfo {author} {\bibfnamefont {L.}~\bibnamefont
  {Pismen}}\ and\ \bibinfo {author} {\bibfnamefont {A.}~\bibnamefont
  {Nepomnyashchy}},\ }\bibfield  {title} {\enquote {\bibinfo {title} {Mobility
  of spiral waves},}\ }\href@noop {} {\bibfield  {journal} {\bibinfo  {journal}
  {Phys. Rev. A}\ }\textbf {\bibinfo {volume} {44}},\ \bibinfo {pages} {R2243}
  (\bibinfo {year} {1991})}\BibitemShut {NoStop}%
\bibitem [{\citenamefont {Pismen}\ and\ \citenamefont
  {Nepomnyashchy}(1992)}]{pismen1992interaction}%
  \BibitemOpen
  \bibfield  {author} {\bibinfo {author} {\bibfnamefont {L.}~\bibnamefont
  {Pismen}}\ and\ \bibinfo {author} {\bibfnamefont {A.}~\bibnamefont
  {Nepomnyashchy}},\ }\bibfield  {title} {\enquote {\bibinfo {title} {On
  interaction of spiral waves},}\ }\href@noop {} {\bibfield  {journal}
  {\bibinfo  {journal} {Physica D}\ }\textbf {\bibinfo {volume} {54}},\
  \bibinfo {pages} {183--193} (\bibinfo {year} {1992})}\BibitemShut {NoStop}%
\bibitem [{\citenamefont {Biktasheva}, \citenamefont {Holden},\ and\
  \citenamefont {Biktashev}(2006)}]{BiHoBi06}%
  \BibitemOpen
  \bibfield  {author} {\bibinfo {author} {\bibfnamefont {I.~V.}\ \bibnamefont
  {Biktasheva}}, \bibinfo {author} {\bibfnamefont {A.~V.}\ \bibnamefont
  {Holden}}, \ and\ \bibinfo {author} {\bibfnamefont {V.~N.}\ \bibnamefont
  {Biktashev}},\ }\bibfield  {title} {\enquote {\bibinfo {title} {Localization
  of response functions of spiral waves in the {F}itz{H}ugh--{N}agumo
  system},}\ }\href@noop {} {\bibfield  {journal} {\bibinfo  {journal} {Int. J.
  Bifur. Chaos}\ }\textbf {\bibinfo {volume} {16}},\ \bibinfo {pages}
  {1547--1555} (\bibinfo {year} {2006})}\BibitemShut {NoStop}%
\bibitem [{\citenamefont {Biktasheva}, \citenamefont {Dierckx},\ and\
  \citenamefont {Biktashev}(2015)}]{biktasheva2015}%
  \BibitemOpen
  \bibfield  {author} {\bibinfo {author} {\bibfnamefont {I.~V.}\ \bibnamefont
  {Biktasheva}}, \bibinfo {author} {\bibfnamefont {H.}~\bibnamefont {Dierckx}},
  \ and\ \bibinfo {author} {\bibfnamefont {V.~N.}\ \bibnamefont {Biktashev}},\
  }\bibfield  {title} {\enquote {\bibinfo {title} {Drift of scroll waves in
  thin layers caused by thickness features: Asymptotic theory and numerical
  simulations},}\ }\href@noop {} {\bibfield  {journal} {\bibinfo  {journal}
  {Phys. Rev. Lett.}\ }\textbf {\bibinfo {volume} {114}},\ \bibinfo {pages}
  {068302} (\bibinfo {year} {2015})}\BibitemShut {NoStop}%
\bibitem [{\citenamefont {Biktashev}, \citenamefont {Biktasheva},\ and\
  \citenamefont {Sarvazyan}(2011)}]{Biktashev2011}%
  \BibitemOpen
  \bibfield  {author} {\bibinfo {author} {\bibfnamefont {V.~N.}\ \bibnamefont
  {Biktashev}}, \bibinfo {author} {\bibfnamefont {I.~V.}\ \bibnamefont
  {Biktasheva}}, \ and\ \bibinfo {author} {\bibfnamefont {N.~A.}\ \bibnamefont
  {Sarvazyan}},\ }\bibfield  {title} {\enquote {\bibinfo {title} {Evolution of
  spiral and scroll waves of excitation in a mathematical model of ischaemic
  border zone},}\ }\href@noop {} {\bibfield  {journal} {\bibinfo  {journal}
  {PLoS One}\ }\textbf {\bibinfo {volume} {6}},\ \bibinfo {pages} {e24388}
  (\bibinfo {year} {2011})}\BibitemShut {NoStop}%
\bibitem [{\citenamefont {Biktashev}(2005)}]{biktashev2005causodynamics}%
  \BibitemOpen
  \bibfield  {author} {\bibinfo {author} {\bibfnamefont {V.}~\bibnamefont
  {Biktashev}},\ }\bibfield  {title} {\enquote {\bibinfo {title} {Causodynamics
  of autowave patterns},}\ }\href@noop {} {\bibfield  {journal} {\bibinfo
  {journal} {Physical review letters}\ }\textbf {\bibinfo {volume} {95}},\
  \bibinfo {pages} {084501} (\bibinfo {year} {2005})}\BibitemShut {NoStop}%
\bibitem [{\citenamefont {Mornev}\ \emph {et~al.}(2003)\citenamefont {Mornev},
  \citenamefont {Tsyganov}, \citenamefont {Aslanidi},\ and\ \citenamefont
  {Tsyganov}}]{Mornev2003}%
  \BibitemOpen
  \bibfield  {author} {\bibinfo {author} {\bibfnamefont {O.~A.}\ \bibnamefont
  {Mornev}}, \bibinfo {author} {\bibfnamefont {I.~M.}\ \bibnamefont
  {Tsyganov}}, \bibinfo {author} {\bibfnamefont {O.~V.}\ \bibnamefont
  {Aslanidi}}, \ and\ \bibinfo {author} {\bibfnamefont {M.~A.}\ \bibnamefont
  {Tsyganov}},\ }\bibfield  {title} {\enquote {\bibinfo {title} {Beyond the
  {K}uramoto-{Z}el{\textquoteright}dovich theory: Steadily rotating concave
  spiral waves and their relation to the echo phenomenon},}\ }\href@noop {}
  {\bibfield  {journal} {\bibinfo  {journal} {Journal of Experimental and
  Theoretical Physics Letters}\ }\textbf {\bibinfo {volume} {77}},\ \bibinfo
  {pages} {270--275} (\bibinfo {year} {2003})}\BibitemShut {NoStop}%
\bibitem [{\citenamefont {Biktasheva}\ and\ \citenamefont
  {Biktashev}(2003)}]{biktasheva2003wave}%
  \BibitemOpen
  \bibfield  {author} {\bibinfo {author} {\bibfnamefont {I.~V.}\ \bibnamefont
  {Biktasheva}}\ and\ \bibinfo {author} {\bibfnamefont {V.~N.}\ \bibnamefont
  {Biktashev}},\ }\bibfield  {title} {\enquote {\bibinfo {title} {Wave-particle
  dualism of spiral waves dynamics},}\ }\href@noop {} {\bibfield  {journal}
  {\bibinfo  {journal} {Phys. Rev. E}\ }\textbf {\bibinfo {volume} {67}},\
  \bibinfo {pages} {026221} (\bibinfo {year} {2003})}\BibitemShut {NoStop}%
\bibitem [{\citenamefont {Langham}\ and\ \citenamefont
  {Barkley}(2013)}]{LanBar13}%
  \BibitemOpen
  \bibfield  {author} {\bibinfo {author} {\bibfnamefont {J.}~\bibnamefont
  {Langham}}\ and\ \bibinfo {author} {\bibfnamefont {D.}~\bibnamefont
  {Barkley}},\ }\bibfield  {title} {\enquote {\bibinfo {title} {Non-specular
  reflections in a macroscopic system with wave-particle duality: {Spiral}
  waves in bounded media},}\ }\href {\doibase 10.1063/1.4793783} {\bibfield
  {journal} {\bibinfo  {journal} {Chaos}\ }\textbf {\bibinfo {volume} {23}},\
  \bibinfo {pages} {013134} (\bibinfo {year} {2013})},\ \bibinfo {note}
  {\arXiv{1304.0591}}\BibitemShut {NoStop}%
\bibitem [{\citenamefont {Langham}, \citenamefont {Biktasheva},\ and\
  \citenamefont {Barkley}(2014)}]{LanBar14}%
  \BibitemOpen
  \bibfield  {author} {\bibinfo {author} {\bibfnamefont {J.}~\bibnamefont
  {Langham}}, \bibinfo {author} {\bibfnamefont {I.}~\bibnamefont {Biktasheva}},
  \ and\ \bibinfo {author} {\bibfnamefont {D.}~\bibnamefont {Barkley}},\
  }\bibfield  {title} {\enquote {\bibinfo {title} {Asymptotic dynamics of
  reflecting spiral waves},}\ }\href@noop {} {\bibfield  {journal} {\bibinfo
  {journal} {Phys. Rev. E}\ }\textbf {\bibinfo {volume} {90}},\ \bibinfo
  {pages} {062902} (\bibinfo {year} {2014})}\BibitemShut {NoStop}%
\bibitem [{\citenamefont {Allexandre}\ and\ \citenamefont
  {Otani}(2004)}]{Otani2004}%
  \BibitemOpen
  \bibfield  {author} {\bibinfo {author} {\bibfnamefont {D.}~\bibnamefont
  {Allexandre}}\ and\ \bibinfo {author} {\bibfnamefont {N.~F.}\ \bibnamefont
  {Otani}},\ }\bibfield  {title} {\enquote {\bibinfo {title} {Preventing
  alternans-induced spiral wave breakup in cardiac tissue: {An}
  ion-channel-based approach},}\ }\href@noop {} {\bibfield  {journal} {\bibinfo
   {journal} {Phys. Rev. E}\ }\textbf {\bibinfo {volume} {70}},\ \bibinfo
  {pages} {061903} (\bibinfo {year} {2004})}\BibitemShut {NoStop}%
\bibitem [{\citenamefont {Fenton}\ and\ \citenamefont
  {Karma}(1998)}]{Fenton1998}%
  \BibitemOpen
  \bibfield  {author} {\bibinfo {author} {\bibfnamefont {F.}~\bibnamefont
  {Fenton}}\ and\ \bibinfo {author} {\bibfnamefont {A.}~\bibnamefont {Karma}},\
  }\bibfield  {title} {\enquote {\bibinfo {title} {Vortex dynamics in
  three-dimensional continuous myocardium with fiber rotation: {Filament}
  instability and fibrillation},}\ }\href@noop {} {\bibfield  {journal}
  {\bibinfo  {journal} {Chaos}\ }\textbf {\bibinfo {volume} {8}},\ \bibinfo
  {pages} {20--47} (\bibinfo {year} {1998})}\BibitemShut {NoStop}%
\bibitem [{\citenamefont {Garz{\'o}n}, \citenamefont {Grigoriev},\ and\
  \citenamefont {Fenton}(2011)}]{Garzon11}%
  \BibitemOpen
  \bibfield  {author} {\bibinfo {author} {\bibfnamefont {A.}~\bibnamefont
  {Garz{\'o}n}}, \bibinfo {author} {\bibfnamefont {R.~O.}\ \bibnamefont
  {Grigoriev}}, \ and\ \bibinfo {author} {\bibfnamefont {F.~H.}\ \bibnamefont
  {Fenton}},\ }\bibfield  {title} {\enquote {\bibinfo {title} {Model-based
  control of cardiac alternans in {Purkinje} fibers},}\ }\href@noop {}
  {\bibfield  {journal} {\bibinfo  {journal} {Phys. Rev. E}\ }\textbf {\bibinfo
  {volume} {84}},\ \bibinfo {pages} {041927} (\bibinfo {year}
  {2011})}\BibitemShut {NoStop}%
\bibitem [{\citenamefont {Garz{\'o}n}, \citenamefont {Grigoriev},\ and\
  \citenamefont {Fenton}(2014)}]{Garzon14}%
  \BibitemOpen
  \bibfield  {author} {\bibinfo {author} {\bibfnamefont {A.}~\bibnamefont
  {Garz{\'o}n}}, \bibinfo {author} {\bibfnamefont {R.~O.}\ \bibnamefont
  {Grigoriev}}, \ and\ \bibinfo {author} {\bibfnamefont {F.~H.}\ \bibnamefont
  {Fenton}},\ }\bibfield  {title} {\enquote {\bibinfo {title} {Continuous-time
  control of alternans in long purkinje fibers},}\ }\href@noop {} {\bibfield
  {journal} {\bibinfo  {journal} {Chaos}\ }\textbf {\bibinfo {volume} {24}},\
  \bibinfo {pages} {033124} (\bibinfo {year} {2014})}\BibitemShut {NoStop}%
\bibitem [{\citenamefont {Karma}(1993)}]{Karma1993}%
  \BibitemOpen
  \bibfield  {author} {\bibinfo {author} {\bibfnamefont {A.}~\bibnamefont
  {Karma}},\ }\bibfield  {title} {\enquote {\bibinfo {title} {Spiral breakup in
  model equations of action-potential propagation in cardiac tissue},}\
  }\href@noop {} {\bibfield  {journal} {\bibinfo  {journal} {Phys. Rev. Lett.}\
  }\textbf {\bibinfo {volume} {71}},\ \bibinfo {pages} {1103--1106} (\bibinfo
  {year} {1993})}\BibitemShut {NoStop}%
\bibitem [{\citenamefont {Karma}(1994)}]{karma94}%
  \BibitemOpen
  \bibfield  {author} {\bibinfo {author} {\bibfnamefont {A.}~\bibnamefont
  {Karma}},\ }\bibfield  {title} {\enquote {\bibinfo {title} {Electrical
  alternans and spiral wave breakup in cardiac tissue},}\ }\href {\doibase
  10.1063/1.166024} {\bibfield  {journal} {\bibinfo  {journal} {Chaos}\
  }\textbf {\bibinfo {volume} {4}},\ \bibinfo {pages} {461--472} (\bibinfo
  {year} {1994})}\BibitemShut {NoStop}%
\bibitem [{\citenamefont {Pastore}\ \emph {et~al.}(1999)\citenamefont
  {Pastore}, \citenamefont {Girouard}, \citenamefont {Laurita}, \citenamefont
  {Akar},\ and\ \citenamefont {Rosenbaum}}]{Pastore1999}%
  \BibitemOpen
  \bibfield  {author} {\bibinfo {author} {\bibfnamefont {J.~M.}\ \bibnamefont
  {Pastore}}, \bibinfo {author} {\bibfnamefont {S.~D.}\ \bibnamefont
  {Girouard}}, \bibinfo {author} {\bibfnamefont {K.~R.}\ \bibnamefont
  {Laurita}}, \bibinfo {author} {\bibfnamefont {F.~G.}\ \bibnamefont {Akar}}, \
  and\ \bibinfo {author} {\bibfnamefont {D.~S.}\ \bibnamefont {Rosenbaum}},\
  }\bibfield  {title} {\enquote {\bibinfo {title} {Mechanism linking {T}-wave
  alternans to the genesis of cardiac fibrillation},}\ }\href@noop {}
  {\bibfield  {journal} {\bibinfo  {journal} {Circulation}\ }\textbf {\bibinfo
  {volume} {99}},\ \bibinfo {pages} {1385--1394} (\bibinfo {year}
  {1999})}\BibitemShut {NoStop}%
\bibitem [{\citenamefont {Jalife}(2000)}]{Jalife2000}%
  \BibitemOpen
  \bibfield  {author} {\bibinfo {author} {\bibfnamefont {J.}~\bibnamefont
  {Jalife}},\ }\bibfield  {title} {\enquote {\bibinfo {title} {Ventricular
  fibrillation: mechanisms of initiation and maintenance},}\ }\href@noop {}
  {\bibfield  {journal} {\bibinfo  {journal} {Annual Rev. Physiol.}\ }\textbf
  {\bibinfo {volume} {62}},\ \bibinfo {pages} {25} (\bibinfo {year}
  {2000})}\BibitemShut {NoStop}%
\bibitem [{\citenamefont {ten Tusscher}\ and\ \citenamefont
  {Panfilov}(2006)}]{ten2006alternans}%
  \BibitemOpen
  \bibfield  {author} {\bibinfo {author} {\bibfnamefont {K.~H.}\ \bibnamefont
  {ten Tusscher}}\ and\ \bibinfo {author} {\bibfnamefont {A.~V.}\ \bibnamefont
  {Panfilov}},\ }\bibfield  {title} {\enquote {\bibinfo {title} {Alternans and
  spiral breakup in a human ventricular tissue model},}\ }\href@noop {}
  {\bibfield  {journal} {\bibinfo  {journal} {American Journal of
  Physiology-Heart and Circulatory Physiology}\ }\textbf {\bibinfo {volume}
  {291}},\ \bibinfo {pages} {H1088--H1100} (\bibinfo {year}
  {2006})}\BibitemShut {NoStop}%
\bibitem [{\citenamefont {Nolasco}\ and\ \citenamefont
  {Dahlen}(1968)}]{Nolasco1968}%
  \BibitemOpen
  \bibfield  {author} {\bibinfo {author} {\bibfnamefont {J.~B.}\ \bibnamefont
  {Nolasco}}\ and\ \bibinfo {author} {\bibfnamefont {R.~W.}\ \bibnamefont
  {Dahlen}},\ }\bibfield  {title} {\enquote {\bibinfo {title} {A graphic method
  for the study of alternation in cardiac action potentials},}\ }\href@noop {}
  {\bibfield  {journal} {\bibinfo  {journal} {J. Appl. Physiol.}\ }\textbf
  {\bibinfo {volume} {25}},\ \bibinfo {pages} {191--196} (\bibinfo {year}
  {1968})}\BibitemShut {NoStop}%
\bibitem [{\citenamefont {Frame}\ and\ \citenamefont
  {Simson}(1988)}]{frame1988oscillations}%
  \BibitemOpen
  \bibfield  {author} {\bibinfo {author} {\bibfnamefont {L.~H.}\ \bibnamefont
  {Frame}}\ and\ \bibinfo {author} {\bibfnamefont {M.~B.}\ \bibnamefont
  {Simson}},\ }\bibfield  {title} {\enquote {\bibinfo {title} {Oscillations of
  conduction, action potential duration, and refractoriness. a mechanism for
  spontaneous termination of reentrant tachycardias.}}\ }\href@noop {}
  {\bibfield  {journal} {\bibinfo  {journal} {Circulation}\ }\textbf {\bibinfo
  {volume} {78}},\ \bibinfo {pages} {1277--1287} (\bibinfo {year}
  {1988})}\BibitemShut {NoStop}%
\bibitem [{\citenamefont {Marcotte}\ and\ \citenamefont
  {Grigoriev}(2015)}]{Marcotte2015}%
  \BibitemOpen
  \bibfield  {author} {\bibinfo {author} {\bibfnamefont {C.~D.}\ \bibnamefont
  {Marcotte}}\ and\ \bibinfo {author} {\bibfnamefont {R.~O.}\ \bibnamefont
  {Grigoriev}},\ }\bibfield  {title} {\enquote {\bibinfo {title} {Unstable
  spiral waves and local {Euclid}ean symmetry in a model of cardiac tissue},}\
  }\href@noop {} {\bibfield  {journal} {\bibinfo  {journal} {Chaos}\ }\textbf
  {\bibinfo {volume} {25}},\ \bibinfo {pages} {063116} (\bibinfo {year}
  {2015})}\BibitemShut {NoStop}%
\bibitem [{\citenamefont {Biktashev}, \citenamefont {Holden},\ and\
  \citenamefont {Nikolaev}(1996)}]{BiHoNi96}%
  \BibitemOpen
  \bibfield  {author} {\bibinfo {author} {\bibfnamefont {V.~N.}\ \bibnamefont
  {Biktashev}}, \bibinfo {author} {\bibfnamefont {A.~V.}\ \bibnamefont
  {Holden}}, \ and\ \bibinfo {author} {\bibfnamefont {E.~V.}\ \bibnamefont
  {Nikolaev}},\ }\bibfield  {title} {\enquote {\bibinfo {title} {Spiral wave
  meander and symmetry of the plane},}\ }\href@noop {} {\bibfield  {journal}
  {\bibinfo  {journal} {Int. J. Bifur. Chaos}\ }\textbf {\bibinfo {volume}
  {6}},\ \bibinfo {pages} {2433--2440} (\bibinfo {year} {1996})}\BibitemShut
  {NoStop}%
\bibitem [{\citenamefont {Barkley}(1992)}]{barkley1992}%
  \BibitemOpen
  \bibfield  {author} {\bibinfo {author} {\bibfnamefont {D.}~\bibnamefont
  {Barkley}},\ }\bibfield  {title} {\enquote {\bibinfo {title} {Linear
  stability analysis of rotating spiral waves in excitable media},}\ }\href
  {\doibase 10.1103/PhysRevLett.68.2090} {\bibfield  {journal} {\bibinfo
  {journal} {Phys. Rev. Lett.}\ }\textbf {\bibinfo {volume} {68}},\ \bibinfo
  {pages} {2090--2093} (\bibinfo {year} {1992})}\BibitemShut {NoStop}%
\bibitem [{\citenamefont {Byrne}, \citenamefont {Marcotte},\ and\ \citenamefont
  {Grigoriev}(2015)}]{ByMaGr14}%
  \BibitemOpen
  \bibfield  {author} {\bibinfo {author} {\bibfnamefont {G.}~\bibnamefont
  {Byrne}}, \bibinfo {author} {\bibfnamefont {C.~D.}\ \bibnamefont {Marcotte}},
  \ and\ \bibinfo {author} {\bibfnamefont {R.~O.}\ \bibnamefont {Grigoriev}},\
  }\bibfield  {title} {\enquote {\bibinfo {title} {Exact coherent structures
  and chaotic dynamics in a model of cardiac tissue},}\ }\href@noop {}
  {\bibfield  {journal} {\bibinfo  {journal} {Chaos}\ }\textbf {\bibinfo
  {volume} {25}},\ \bibinfo {pages} {033108} (\bibinfo {year}
  {2015})}\BibitemShut {NoStop}%
\bibitem [{\citenamefont {Arnoldi}(1951)}]{Arnoldi1951}%
  \BibitemOpen
  \bibfield  {author} {\bibinfo {author} {\bibfnamefont {W.~E.}\ \bibnamefont
  {Arnoldi}},\ }\bibfield  {title} {\enquote {\bibinfo {title} {The principle
  of minimized iterations in the solution of the matrix eigenvalue problem},}\
  }\href@noop {} {\bibfield  {journal} {\bibinfo  {journal} {Quart. Appl.
  Math.}\ }\textbf {\bibinfo {volume} {9}},\ \bibinfo {pages} {17--29}
  (\bibinfo {year} {1951})}\BibitemShut {NoStop}%
\bibitem [{\citenamefont {Saad}\ and\ \citenamefont
  {Schultz}(1986)}]{saad1986gmres}%
  \BibitemOpen
  \bibfield  {author} {\bibinfo {author} {\bibfnamefont {Y.}~\bibnamefont
  {Saad}}\ and\ \bibinfo {author} {\bibfnamefont {M.~H.}\ \bibnamefont
  {Schultz}},\ }\bibfield  {title} {\enquote {\bibinfo {title} {{GM}res: A
  generalized minimal residual algorithm for solving nonsymmetric linear
  systems},}\ }\href@noop {} {\bibfield  {journal} {\bibinfo  {journal} {SIAM
  J. Sci. Stat. Comp.}\ }\textbf {\bibinfo {volume} {7}},\ \bibinfo {pages}
  {856--869} (\bibinfo {year} {1986})}\BibitemShut {NoStop}%
\bibitem [{\citenamefont {Biktasheva}\ \emph {et~al.}(2009)\citenamefont
  {Biktasheva}, \citenamefont {Barkley}, \citenamefont {Biktashev},
  \citenamefont {Bordyugov},\ and\ \citenamefont {Foulkes}}]{BBBBF09}%
  \BibitemOpen
  \bibfield  {author} {\bibinfo {author} {\bibfnamefont {I.~V.}\ \bibnamefont
  {Biktasheva}}, \bibinfo {author} {\bibfnamefont {D.}~\bibnamefont {Barkley}},
  \bibinfo {author} {\bibfnamefont {V.~N.}\ \bibnamefont {Biktashev}}, \bibinfo
  {author} {\bibfnamefont {G.~V.}\ \bibnamefont {Bordyugov}}, \ and\ \bibinfo
  {author} {\bibfnamefont {A.~J.}\ \bibnamefont {Foulkes}},\ }\bibfield
  {title} {\enquote {\bibinfo {title} {Computation of the response functions of
  spiral waves in active media},}\ }\href@noop {} {\bibfield  {journal}
  {\bibinfo  {journal} {Phys. Rev. E}\ }\textbf {\bibinfo {volume} {79}},\
  \bibinfo {pages} {056702} (\bibinfo {year} {2009})}\BibitemShut {NoStop}%
\bibitem [{\citenamefont {Sandstede}\ and\ \citenamefont
  {Scheel}(2000{\natexlab{a}})}]{Sandstede:2000ab}%
  \BibitemOpen
  \bibfield  {author} {\bibinfo {author} {\bibfnamefont {B.}~\bibnamefont
  {Sandstede}}\ and\ \bibinfo {author} {\bibfnamefont {A.}~\bibnamefont
  {Scheel}},\ }\bibfield  {title} {\enquote {\bibinfo {title} {Absolute and
  convective instabilities of waves on unbounded and large bounded domains},}\
  }\href@noop {} {\bibfield  {journal} {\bibinfo  {journal} {Physica D}\
  }\textbf {\bibinfo {volume} {145}},\ \bibinfo {pages} {233--277} (\bibinfo
  {year} {2000}{\natexlab{a}})}\BibitemShut {NoStop}%
\bibitem [{\citenamefont {Sandstede}\ and\ \citenamefont
  {Scheel}(2000{\natexlab{b}})}]{Sandstede:2000abs}%
  \BibitemOpen
  \bibfield  {author} {\bibinfo {author} {\bibfnamefont {B.}~\bibnamefont
  {Sandstede}}\ and\ \bibinfo {author} {\bibfnamefont {A.}~\bibnamefont
  {Scheel}},\ }\bibfield  {title} {\enquote {\bibinfo {title} {Absolute versus
  convective instability of spiral waves},}\ }\href@noop {} {\bibfield
  {journal} {\bibinfo  {journal} {Phys. Rev. E}\ }\textbf {\bibinfo {volume}
  {62}},\ \bibinfo {pages} {7708--7714} (\bibinfo {year}
  {2000}{\natexlab{b}})}\BibitemShut {NoStop}%
\bibitem [{\citenamefont {Wheeler}\ and\ \citenamefont
  {Barkley}(2006)}]{Wheeler:2006co}%
  \BibitemOpen
  \bibfield  {author} {\bibinfo {author} {\bibfnamefont {P.}~\bibnamefont
  {Wheeler}}\ and\ \bibinfo {author} {\bibfnamefont {D.}~\bibnamefont
  {Barkley}},\ }\bibfield  {title} {\enquote {\bibinfo {title} {Computation of
  spiral spectra},}\ }\href@noop {} {\bibfield  {journal} {\bibinfo  {journal}
  {SIAM J. Appl. Dyn. Syst.}\ } (\bibinfo {year} {2006})}\BibitemShut {NoStop}%
\bibitem [{\citenamefont {Sandstede}\ and\ \citenamefont
  {Scheel}(2004)}]{sandstede2004defects}%
  \BibitemOpen
  \bibfield  {author} {\bibinfo {author} {\bibfnamefont {B.}~\bibnamefont
  {Sandstede}}\ and\ \bibinfo {author} {\bibfnamefont {A.}~\bibnamefont
  {Scheel}},\ }\bibfield  {title} {\enquote {\bibinfo {title} {Defects in
  oscillatory media: toward a classification},}\ }\href@noop {} {\bibfield
  {journal} {\bibinfo  {journal} {SIAM Journal on Applied Dynamical Systems}\
  }\textbf {\bibinfo {volume} {3}},\ \bibinfo {pages} {1--68} (\bibinfo {year}
  {2004})}\BibitemShut {NoStop}%
\bibitem [{\citenamefont {Biktasheva}, \citenamefont {Biktashev},\ and\
  \citenamefont {Foulkes}(2010)}]{Biktasheva:2010co}%
  \BibitemOpen
  \bibfield  {author} {\bibinfo {author} {\bibfnamefont {I.~V.}\ \bibnamefont
  {Biktasheva}}, \bibinfo {author} {\bibfnamefont {V.~N.}\ \bibnamefont
  {Biktashev}}, \ and\ \bibinfo {author} {\bibfnamefont {A.~J.}\ \bibnamefont
  {Foulkes}},\ }\bibfield  {title} {\enquote {\bibinfo {title} {Computation of
  the drift velocity of spiral waves using response functions},}\ }\href@noop
  {} {\bibfield  {journal} {\bibinfo  {journal} {Phys. Rev. E}\ }\textbf
  {\bibinfo {volume} {81}},\ \bibinfo {pages} {066202} (\bibinfo {year}
  {2010})}\BibitemShut {NoStop}%
\bibitem [{\citenamefont {Aranson}, \citenamefont {Kramer},\ and\ \citenamefont
  {Weber}(1993)}]{aranson1993interaction}%
  \BibitemOpen
  \bibfield  {author} {\bibinfo {author} {\bibfnamefont {I.~S.}\ \bibnamefont
  {Aranson}}, \bibinfo {author} {\bibfnamefont {L.}~\bibnamefont {Kramer}}, \
  and\ \bibinfo {author} {\bibfnamefont {A.}~\bibnamefont {Weber}},\ }\bibfield
   {title} {\enquote {\bibinfo {title} {Theory of interaction and bound states
  of spiral waves in oscillatory media},}\ }\href {\doibase
  10.1103/PhysRevE.47.3231} {\bibfield  {journal} {\bibinfo  {journal} {Phys.
  Rev. E}\ }\textbf {\bibinfo {volume} {47}},\ \bibinfo {pages} {3231--3241}
  (\bibinfo {year} {1993})}\BibitemShut {NoStop}%
\bibitem [{\citenamefont {Aranson}, \citenamefont {Kessler},\ and\
  \citenamefont {Mitkov}(1995)}]{aranson1995drift}%
  \BibitemOpen
  \bibfield  {author} {\bibinfo {author} {\bibfnamefont {I.}~\bibnamefont
  {Aranson}}, \bibinfo {author} {\bibfnamefont {D.}~\bibnamefont {Kessler}}, \
  and\ \bibinfo {author} {\bibfnamefont {I.}~\bibnamefont {Mitkov}},\
  }\bibfield  {title} {\enquote {\bibinfo {title} {Drift of spiral waves in
  excitable media},}\ }\href@noop {} {\bibfield  {journal} {\bibinfo  {journal}
  {Physica D}\ }\textbf {\bibinfo {volume} {85}},\ \bibinfo {pages} {142--155}
  (\bibinfo {year} {1995})}\BibitemShut {NoStop}%
\bibitem [{\citenamefont {Biktashev}\ and\ \citenamefont
  {Biktasheva}(2010)}]{Biktashev2010}%
  \BibitemOpen
  \bibfield  {author} {\bibinfo {author} {\bibfnamefont {V.~N.}\ \bibnamefont
  {Biktashev}}\ and\ \bibinfo {author} {\bibfnamefont {I.~V.}\ \bibnamefont
  {Biktasheva}},\ }\bibfield  {title} {\enquote {\bibinfo {title} {Orbital
  motion of spiral waves in excitable media},}\ }\href@noop {} {\bibfield
  {journal} {\bibinfo  {journal} {Phys. Rev. Lett.}\ }\textbf {\bibinfo
  {volume} {104}},\ \bibinfo {pages} {058302} (\bibinfo {year}
  {2010})}\BibitemShut {NoStop}%
\bibitem [{\citenamefont {Biktasheva}(2000)}]{biktasheva2000drift}%
  \BibitemOpen
  \bibfield  {author} {\bibinfo {author} {\bibfnamefont {I.}~\bibnamefont
  {Biktasheva}},\ }\bibfield  {title} {\enquote {\bibinfo {title} {Drift of
  spiral waves in the complex {Ginzburg-Landau} equation due to media
  inhomogeneities},}\ }\href@noop {} {\bibfield  {journal} {\bibinfo  {journal}
  {Phys. Rev. E}\ }\textbf {\bibinfo {volume} {62}},\ \bibinfo {pages} {8800}
  (\bibinfo {year} {2000})}\BibitemShut {NoStop}%
\bibitem [{\citenamefont {Marcotte}\ and\ \citenamefont
  {Grigoriev}(2016)}]{Marcotte2016b}%
  \BibitemOpen
  \bibfield  {author} {\bibinfo {author} {\bibfnamefont {C.~D.}\ \bibnamefont
  {Marcotte}}\ and\ \bibinfo {author} {\bibfnamefont {R.~O.}\ \bibnamefont
  {Grigoriev}},\ }\href@noop {} {\enquote {\bibinfo {title} {Adjoint
  eigenfunctions of temporally recurrent multi-spiral solutions in a simple
  model of atrial fibrillation.}}\ } (\bibinfo {year} {2016}),\ \bibinfo {note}
  {\arxiv{1605.00115}}\BibitemShut {NoStop}%
\bibitem [{\citenamefont {Agladze}, \citenamefont {Davydov},\ and\
  \citenamefont {Mikhailov}(1987)}]{agladze1987}%
  \BibitemOpen
  \bibfield  {author} {\bibinfo {author} {\bibfnamefont {K.~I.}\ \bibnamefont
  {Agladze}}, \bibinfo {author} {\bibfnamefont {V.~A.}\ \bibnamefont
  {Davydov}}, \ and\ \bibinfo {author} {\bibfnamefont {A.~S.}\ \bibnamefont
  {Mikhailov}},\ }\bibfield  {title} {\enquote {\bibinfo {title} {An
  observation of resonance of spiral waves in distributed excitable medium},}\
  }\href@noop {} {\bibfield  {journal} {\bibinfo  {journal} {JETP Lett}\
  }\textbf {\bibinfo {volume} {45}},\ \bibinfo {pages} {767--770} (\bibinfo
  {year} {1987})}\BibitemShut {NoStop}%
\bibitem [{\citenamefont {Efimov}, \citenamefont {Gray},\ and\ \citenamefont
  {Roth}(2000)}]{efimov2000}%
  \BibitemOpen
  \bibfield  {author} {\bibinfo {author} {\bibfnamefont {I.~R.}\ \bibnamefont
  {Efimov}}, \bibinfo {author} {\bibfnamefont {R.~A.}\ \bibnamefont {Gray}}, \
  and\ \bibinfo {author} {\bibfnamefont {B.~J.}\ \bibnamefont {Roth}},\
  }\bibfield  {title} {\enquote {\bibinfo {title} {Virtual electrodes and
  deexcitation: new insights into fibrillation induction and defibrillation},}\
  }\href@noop {} {\bibfield  {journal} {\bibinfo  {journal} {Journal of
  cardiovascular electrophysiology}\ }\textbf {\bibinfo {volume} {11}},\
  \bibinfo {pages} {339--353} (\bibinfo {year} {2000})}\BibitemShut {NoStop}%
\bibitem [{\citenamefont {Fenton}\ \emph {et~al.}(2009)\citenamefont {Fenton},
  \citenamefont {Luther}, \citenamefont {Cherry}, \citenamefont {Otani},
  \citenamefont {Krinksy}, \citenamefont {Pumir}, \citenamefont {Bodenschatz},\
  and\ \citenamefont {Jr.}}]{Fenton2009}%
  \BibitemOpen
  \bibfield  {author} {\bibinfo {author} {\bibfnamefont {F.~H.}\ \bibnamefont
  {Fenton}}, \bibinfo {author} {\bibfnamefont {S.}~\bibnamefont {Luther}},
  \bibinfo {author} {\bibfnamefont {E.~M.}\ \bibnamefont {Cherry}}, \bibinfo
  {author} {\bibfnamefont {N.~F.}\ \bibnamefont {Otani}}, \bibinfo {author}
  {\bibfnamefont {V.}~\bibnamefont {Krinksy}}, \bibinfo {author} {\bibfnamefont
  {A.}~\bibnamefont {Pumir}}, \bibinfo {author} {\bibfnamefont
  {E.}~\bibnamefont {Bodenschatz}}, \ and\ \bibinfo {author} {\bibfnamefont
  {R.~F.~G.}\ \bibnamefont {Jr.}},\ }\bibfield  {title} {\enquote {\bibinfo
  {title} {Termination of atrial fibrillation using pulsed low-energy far-field
  stimulation},}\ }\href@noop {} {\bibfield  {journal} {\bibinfo  {journal}
  {Circulation}\ }\textbf {\bibinfo {volume} {120}},\ \bibinfo {pages}
  {467--476} (\bibinfo {year} {2009})}\BibitemShut {NoStop}%
\bibitem [{\citenamefont {Hager}(2000)}]{hager2000runge}%
  \BibitemOpen
  \bibfield  {author} {\bibinfo {author} {\bibfnamefont {W.~W.}\ \bibnamefont
  {Hager}},\ }\bibfield  {title} {\enquote {\bibinfo {title} {Runge-kutta
  methods in optimal control and the transformed adjoint system},}\ }\href
  {\doibase 10.1007/s002110000178} {\bibfield  {journal} {\bibinfo  {journal}
  {Numerische Mathematik}\ }\textbf {\bibinfo {volume} {87}},\ \bibinfo {pages}
  {247--282} (\bibinfo {year} {2000})}\BibitemShut {NoStop}%
\bibitem [{\citenamefont {Sandu}(2006)}]{sandu2006properties}%
  \BibitemOpen
  \bibfield  {author} {\bibinfo {author} {\bibfnamefont {A.}~\bibnamefont
  {Sandu}},\ }\bibfield  {title} {\enquote {\bibinfo {title} {On the properties
  of runge-kutta discrete adjoints},}\ }in\ \href@noop {} {\emph {\bibinfo
  {booktitle} {Computational Science--ICCS 2006}}}\ (\bibinfo  {publisher}
  {Springer},\ \bibinfo {year} {2006})\ pp.\ \bibinfo {pages}
  {550--557}\BibitemShut {NoStop}%
\bibitem [{\citenamefont {Enright}\ \emph {et~al.}(1986)\citenamefont
  {Enright}, \citenamefont {Jackson}, \citenamefont {N{\o}rsett},\ and\
  \citenamefont {Thomsen}}]{enright1986interpolants}%
  \BibitemOpen
  \bibfield  {author} {\bibinfo {author} {\bibfnamefont {W.~H.}\ \bibnamefont
  {Enright}}, \bibinfo {author} {\bibfnamefont {K.}~\bibnamefont {Jackson}},
  \bibinfo {author} {\bibfnamefont {S.~P.}\ \bibnamefont {N{\o}rsett}}, \ and\
  \bibinfo {author} {\bibfnamefont {P.~G.}\ \bibnamefont {Thomsen}},\
  }\bibfield  {title} {\enquote {\bibinfo {title} {Interpolants for runge-kutta
  formulas},}\ }\href@noop {} {\bibfield  {journal} {\bibinfo  {journal} {ACM
  Transactions on Mathematical Software (TOMS)}\ }\textbf {\bibinfo {volume}
  {12}},\ \bibinfo {pages} {193--218} (\bibinfo {year} {1986})}\BibitemShut
  {NoStop}%
\end{thebibliography}%

\end{document}